\documentclass[zpreprint,zbstepj]{zeus_paper}
\usepackage[english]{babel}
\usepackage{hangcaption}
\usepackage{multirow}
\chardef\usc=95
\chardef\til=126
\catcode`\@=11 
\DeclareRobustCommand\xdotspace{\futurelet\@let@token\@xdotspace}
\def\@xdotspace{%
  \ifx\@let@token.\else
  \ifx\@let@token\bgroup.\else
  \ifx\@let@token\egroup.\else
  \ifx\@let@token\/.\else
  \ifx\@let@token\ .\else
  \ifx\@let@token~.\else
  \ifx\@let@token!.\else
  \ifx\@let@token,.\else
  \ifx\@let@token:.\else
  \ifx\@let@token;.\else
  \ifx\@let@token?.\else
  \ifx\@let@token/.\else
  \ifx\@let@token'.\else
  \ifx\@let@token).\else
  \ifx\@let@token-.\else
  \ifx\@let@token\@xobeysp.\else
  \ifx\@let@token\space.\else
  \ifx\@let@token\@sptoken.\else
   .\space
   \fi\fi\fi\fi\fi\fi\fi\fi\fi\fi\fi\fi\fi\fi\fi\fi\fi\fi}
\catcode`\@=12 

\newcommand{\stru}[2]{%
   \relax\ifmmode\hbox{\vrule height#1 depth#2 width0pt}%
   \else\vrule height#1 depth#2 width0pt\fi}

\newcommand{\Ronum}[1]{\uppercase\expandafter{\romannumeral#1}}
\newcommand{\ronum}[1]{\expandafter{\romannumeral#1}}
\DeclareRobustCommand{\LaTeXZ}{%
  \LaTeX\kern-.05em4\kern-.1em
  {\raisebox{-0.2ex}{$\scriptstyle\text{ZEUS}$}}\xspace}

\DeclareMathAlphabet{\mathbf}{OT1}{cmr}{bx}{sl}
\newcommand{\eVdist}{\kern-0.06667em}

\newcommand{\Gev}{{\text{Ge}\eVdist\text{V\/}}}

\newcommand{\gev}{{\,\text{Ge}\eVdist\text{V\/}}}

\newcommand{\nb}{\,\text{nb}}

\newcommand{\pbi}{\,\text{pb}^{-1}}

\newcommand{\slashfrac}[2]{%
  \raisebox{0.5ex}{\ensuremath #1}\kern-0.12em/\kern-0.08em
  \raisebox{-.8ex}{\ensuremath #2}}

\newcommand{\sqr}[3]{%
    {\vcenter{\hrule height.#3ex\hbox{\vrule width.#2ex height#1ex
     \kern#1ex\vrule width.#3ex}\hrule height.#2ex}}}

\catcode`\@=11 
\newcommand{\parenbar}{\mathpalette\p@renb@r}
\def\p@renb@r#1#2{\vbox{%
  \ifx#1\scriptscriptstyle \dimen@.7em\dimen@ii.2em\else
  \ifx#1\scriptstyle \dimen@.8em\dimen@ii.25em\else
  \dimen@1em\dimen@ii.4em\fi\fi \offinterlineskip
  \ialign{\hfill##\hfill\cr
    \vbox{\hrule width\dimen@ii}\cr
    \noalign{\vskip-.3ex}%
    \hbox to\dimen@{$\mathchar300\hfil\mathchar301$}\cr
    \noalign{\vskip-.3ex}%
    $#1#2$\cr}}}
\catcode`\@=12 

\newcommand{\IP}{{\rm I$\kern-0.01667em$P}\xspace}

\mathchardef\qsm=63
\mathchardef\pls=43
\mathchardef\mns=512
\mathchardef\plm=518
\mathchardef\eql=61
\mathchardef\smallleft=300
\mathchardef\smallright=301
\mathchardef\les=316
\mathchardef\gre=318
\mathchardef\leq=532
\mathchardef\grq=533

\catcode`\@=11 
\newcounter{pict@width}
\newcounter{pict@height}
\newlength{\pict@scale}
\newcommand{\psfigadd}[4]{%
\setcounter{pict@width}{1*\ratio{#2+\pict@scale/2}{\pict@scale}}
\setcounter{pict@height}{1*\ratio{#3+\pict@scale/2}{\pict@scale}}
\setlength{\unitlength}{\pict@scale}
\hbox to #2{\hspace{-\fill}\begin{picture}(\thepict@width,\thepict@height)
\put(0,0){\psfig{figure=#1,width=#2,height=#3,clip=}}
\SetScale{0.283466457}
\SetWidth{1.763889}
{#4}
\end{picture}}
}
\newcounter{pict@widthfst}
\newcounter{pict@widthscd}
\newcounter{pict@widthtot}
\newcommand{\psfigaddtwo}[7]{%
\setcounter{pict@widthfst}{1*\ratio{#2+\pict@scale/2}{\pict@scale}}
\setcounter{pict@widthscd}{1*\ratio{#2+#4+\pict@scale/2}{\pict@scale}}
\setcounter{pict@widthtot}{1*\ratio{#2+#4+#6+\pict@scale/2}{\pict@scale}}
\setcounter{pict@height}{1*\ratio{#3+\pict@scale/2}{\pict@scale}}
\setlength{\unitlength}{\pict@scale}
\hbox{\hspace{-\fill}\begin{picture}(\thepict@widthtot,\thepict@height)
\put(0,0){\psfig{figure=#1,width=#2,height=#3,clip=}}
\put(\thepict@widthscd,0){\psfig{figure=#5,width=#6,height=#3,clip=}}
\SetScale{0.283466457}
\SetWidth{1.763889}
{#7}
\end{picture}}
}
\newcommand{\psfigror}[4]{%
\setcounter{pict@width}{1*\ratio{#2+\pict@scale/2}{\pict@scale}}
\setcounter{pict@height}{1*\ratio{#3+\pict@scale/2}{\pict@scale}}
\setlength{\unitlength}{\pict@scale}
\hbox{\begin{picture}(\thepict@width,\thepict@height)
\put(0,\thepict@height){\psfig{figure=#1,width=#3,height=#2,clip=,angle=270}}
\SetScale{0.283466457}
\SetWidth{1.763889}
{#4}
\end{picture}}
}
\newcommand{\psfigrol}[4]{%
\setcounter{pict@width}{1*\ratio{#2+\pict@scale/2}{\pict@scale}}
\setcounter{pict@height}{1*\ratio{#3+\pict@scale/2}{\pict@scale}}
\setlength{\unitlength}{\pict@scale}
\hbox{\begin{picture}(\thepict@width,\thepict@height)
\put(0,0){\psfig{figure=#1,width=#3,height=#2,clip=,angle=90}}
\SetScale{0.283466457}
\SetWidth{1.763889}
{#4}
\end{picture}}
}
\catcode`\@=12 
\newlength\listtextwidth

\newcommand{\pcite}[1]{{\protect\cite{#1}}}

\catcode`\@=11 
\newlength{\@tabfninsert}
\newlength{\@tabfnwidth}
\newcommand{\tabfootnote}[2]{%
  \setlength{\@tabfninsert}{0.8em}
  \setlength{\@tabfnwidth}{\textwidth}
  \addtolength{\@tabfnwidth}{-\@tabfninsert}
  \addtolength{\@tabfnwidth}{-0.4em}
  \noindent\makebox[\@tabfninsert][r]{\footnotesize$^{#1}$\hfil}\hfill%
  \parbox[t]{\@tabfnwidth}{\footnotesize #2\hfill}}
\catcode`\@=12 

\def\citeCTD{{\cite{%
nim:a279:290,*npps:b32:181,*nim:a338:254%
}}\xspace}
\def\citeCAL{{\cite{%
nim:a309:77,*nim:a309:101,*nim:a321:356,*nim:a336:23%
}}\xspace}

\newcommand{\coll}{Coll.\xspace}
\newcommand{\etal}{et al.\xspace}
\newcommand {\rzfzz}     {\mbox{${r^{04}_{00}}$}}

\newcommand {\rzfpm}     {\mbox{${r^{04}_{1-1}}$}}   
\newcommand {\pom} {I\hspace{-0.2em}P}

\newcommand {\alphapom} {\mbox{$\alpha_{\pom}$}}
\newcommand {\alphappom} {\mbox{$\alpha^\prime_{\pom}$}}

\newcommand {\Wgp}     {\mbox{${W}$}}

\newcommand {\mumu} {\mbox{$\mu^+\mu^-$}}            
            
\newcommand {\Thel}     {\mbox{${\theta_{h}}$}}
\newcommand {\fhel}     {\mbox{${\phi_{h}}$}}
\newcommand{\lap}{\ensuremath{\stackrel{_{\scriptstyle <}}{_{\scriptstyle\sim}}}
}
\newcommand{\gap}{\ensuremath{\stackrel{_{\scriptstyle >}}{_{\scriptstyle\sim}}}
}

\begin{document}
\title{
Exclusive photoproduction of $\mathbf{J/\psi}$ mesons at HERA
}                                                       
                    
\author{ZEUS Collaboration}

\abstract{
The exclusive photoproduction of $J/\psi$ mesons,
$\gamma\, p\, \rightarrow\, J/\psi\, p$, 
has been studied in $ep$ collisions with the ZEUS detector at HERA, 
in the kinematic range $20<\Wgp<290$ GeV, where $\Wgp$ is
the photon-proton centre-of-mass energy.
The $J/\psi$ mesons were reconstructed in the muon and the electron 
decay channels 
using integrated luminosities of 38$\pbi$ and 55$\pbi$, respectively.
The helicity structure of $J/\psi$ production shows that
the hypothesis of $s$-channel helicity conservation is satisfied
within two standard deviations. 
%
%
The total cross section and 
the differential cross-section $d\sigma/dt$,
where $t$ is the squared four-momentum transfer at the proton vertex,
are presented as a function of $W$, for $|t|<1.8$ GeV$^2$.
The $t$ distribution exhibits an exponential shape with a
slope parameter increasing logarithmically with $W$ with a value
$b=4.15 \pm 0.05 (stat.)^{+0.30}_{-0.18} (syst.)\gev^{-2}$ at $W=90\gev$. 
The effective parameters of the Pomeron trajectory are 
$\alphapom(0) = 1.200 \pm 0.009(stat.)^{+0.004}_{-0.010}(syst.)$
and
$\alphappom= 0.115 \pm 0.018(stat.)^{+0.008}_{-0.015}(syst.)~ {\rm GeV}^{-2}$.
}
\makezeustitle

\pagenumbering{Roman}                                                                              
                                                   %
\begin{center}                                                                                     
{                      \Large  The ZEUS Collaboration              }                               
\end{center}                                                                                       
  S.~Chekanov,                                                                                     
  D.~Krakauer,                                                                                     
  S.~Magill,                                                                                       
  B.~Musgrave,                                                                                     
  A.~Pellegrino,                                                                                   
  J.~Repond,                                                                                       
  R.~Yoshida\\                                                                                     
 {\it Argonne National Laboratory, Argonne, Illinois 60439-4815}~$^{n}$                            
\par \filbreak                                                                                     
  M.C.K.~Mattingly \\                                                                              
 {\it Andrews University, Berrien Springs, Michigan 49104-0380}                                    
\par \filbreak                                                                                     
  P.~Antonioli,                                                                                    
  G.~Bari,                                                                                         
  M.~Basile,                                                                                       
  L.~Bellagamba,                                                                                   
  D.~Boscherini,                                                                                   
  A.~Bruni,                                                                                        
  G.~Bruni,                                                                                        
  G.~Cara~Romeo,                                                                                   
  L.~Cifarelli,                                                                                    
  F.~Cindolo,                                                                                      
  A.~Contin,                                                                                       
  M.~Corradi,                                                                                      
  S.~De~Pasquale,                                                                                  
  P.~Giusti,                                                                                       
  G.~Iacobucci,                                                                                    
  G.~Levi,                                                                                         
  A.~Margotti,                                                                                     
  T.~Massam,                                                                                       
  R.~Nania,                                                                                        
  F.~Palmonari,                                                                                    
  A.~Pesci,                                                                                        
  G.~Sartorelli,                                                                                   
  A.~Zichichi  \\                                                                                  
  {\it University and INFN Bologna, Bologna, Italy}~$^{e}$
% ...
\newpage                                                                                           
% ...
$^{\    2}$ on leave of absence at University of                                                   
Erlangen-N\"urnberg, Germany\\                                                                     
% ...
                                                           %
                                                           %
\newpage   
                                                           %
                                                           %
\begin{tabular}[h]{rp{14cm}}                                                                       
$^{a}$ &  supported by the Natural Sciences and Engineering Research                               
          Council of Canada (NSERC) \\                                                             
$^{b}$ &  supported by the German Federal Ministry for Education and                               
          Research (BMBF), under contract numbers HZ1GUA 2, HZ1GUB 0, HZ1PDA 5, HZ1VFA 5\\         
$^{c}$ &  supported by the MINERVA Gesellschaft f\"ur Forschung GmbH, the                          
          Israel Science Foundation, the U.S.-Israel Binational Science                            
          Foundation, the Israel Ministry of Science and the Benozyio Center                       
          for High Energy Physics\\                                                                
$^{d}$ &  supported by the German-Israeli Foundation, the Israel Science                           
          Foundation, and by the Israel Ministry of Science\\                                      
$^{e}$ &  supported by the Italian National Institute for Nuclear Physics (INFN) \\                
$^{f}$ &  supported by the Japanese Ministry of Education, Science and                             
          Culture (the Monbusho) and its grants for Scientific Research\\                          
$^{g}$ &  supported by the Korean Ministry of Education and Korea Science                          
          and Engineering Foundation\\                                                             
$^{h}$ &  supported by the Netherlands Foundation for Research on Matter (FOM)\\                   
$^{i}$ &  supported by the Polish State Committee for Scientific Research,                         
          grant no. 115/E-343/SPUB-M/DESY/P-03/DZ 121/2001-2002\\                                  
$^{j}$ &  partially supported by the German Federal Ministry for Education                         
          and Research (BMBF)\\                                                                    
$^{k}$ &  supported by the Fund for Fundamental Research of Russian Ministry                       
          for Science and Edu\-cation and by the German Federal Ministry for                       
          Education and Research (BMBF)\\                                                          
$^{l}$ &  supported by the Spanish Ministry of Education and Science                               
          through funds provided by CICYT\\                                                        
$^{m}$ &  supported by the Particle Physics and Astronomy Research Council, UK\\                   
$^{n}$ &  supported by the US Department of Energy\\                                               
$^{o}$ &  supported by the US National Science Foundation                                          
\end{tabular}                                                                                      
                                                           %
                                                           %
\clearpage
\pagenumbering{arabic}
\pagestyle{plain}
\section{Introduction}
\label{sec-int}

Exclusive $J/\psi$ photoproduction is expected to
be described by models based on perturbative QCD (pQCD),
since the mass of the charm quark provides a hard 
scale~\cite{zfp:c57:89,*ryskin2,bfgms,bartels,ginzburg,
MRT99,jhep:103:45,pl:b503:277,
np:b603:427,hep-ph-0111391}. 
In such models, the photon fluctuates into a $c\bar{c}$ pair which subsequently
interacts with the proton. 
This interaction is modelled by the exchange of a gluon ladder and 
the cross section is proportional to the square of the gluon density. 
These models predict a rapid rise in the cross section with 
$\Wgp$,
where $\Wgp$ is the photon-proton centre-of-mass energy,
which is caused by the fast increase 
of the gluon density in the proton at the small values of Bjorken $x$. 

Within the framework of Regge phenomenology~\cite{collins},
diffractive interactions at large centre-of-mass energies are 
the result of the $t$-channel exchange of the Pomeron trajectory, 
$\alphapom(t)$, carrying
the quantum numbers of the vacuum. 
The differential cross section at high energies is expressed as
\begin{equation}
\frac{d\sigma}{dt} \propto F(t) \cdot W^{ 4 \cdot [\alphapom(t)-1]},
\label{eq:regge1}
\end{equation}
where $t$ is the squared four-momentum transfer at the proton vertex
and $F(t)$ is a function of $t$ only.
If $d\sigma/dt$ decreases exponentially and the trajectory is linear in $t$,
$\alphapom(t) = \alphapom(0) + \alphappom t$,
the cross section can be expressed as
\begin{equation}
\frac{d\sigma}{dt}   \propto W^{ 4 [\alphapom (0)-1] } \cdot 
                   e^{b(W)t},
\label{eq:regge2}
\end{equation}
where the slope parameter is $b(\Wgp) = b_0 + 4 \alphappom \ln (\Wgp/W_0)$
and $W_0$ is an arbitrary energy scale parameter.
A fit to hadronic data~\cite{pr:d10:170} 
yields the soft-Pomeron parameter $\alphappom = 0.25 \gev^{-2}$.
The pQCD models predict the effective $\alphappom$ 
in the perturbative regime~\cite{bfgms,shep:11:51,jhep:103:45,pl:b366:337}
to be much smaller than $0.25\gev^2$.

Studies of the exclusive, diffractive photoproduction of vector 
charmonium states, i.e. $\gamma\, p\, \rightarrow\, J/\psi\,p$,
at HERA show that the total cross section~\cite{ZEUS971,H1001}
rises steeply with $\Wgp$. In addition,
the $t$ dependence of the cross section
can be fitted by a single exponential, $d\sigma/dt 
\sim e^{-b|t|}$, with $b \sim$ 4.6 GeV$^{-2}$~\cite{ZEUS971,H1001}. 
There are indications that the
slope parameter $b$ has little variation with 
$\Wgp$~\cite{noshrink,H1001}, i.e. little ``shrinkage'' is observed.

In this paper, more precise measurements of the $t$ and $\Wgp$ dependence 
of the exclusive photoproduction cross section $J/\psi$ mesons
are made using the reaction 
$e\, p\, \rightarrow\, e\,\gamma\,p \rightarrow e\,J/\psi\,p$ 
for values of the photon virtuality, $Q^2$, close to zero
in the range $20<\Wgp<290 \gev$. 
With respect to the previous ZEUS result~\cite{ZEUS971}, 
this analysis covers wider ranges in $W$ and $t$ and
has a large increase in statistics, combined with an improved
understanding of the detector and of the background subtraction.
The  cross section for $\gamma\, p\, \rightarrow\, J/\psi\,p$   
and the slope of the differential cross-section 
$d\sigma_{\gamma p \rightarrow J/\psi p}/dt$ 
are studied as a function of $\Wgp$. The Pomeron trajectory
parameters $\alphapom (0)$  and 
$\alphappom$, which describe the $\Wgp$ dependence of the cross
section and the shrinkage,
respectively, are determined.
The helicity structure of $J/\psi$ production is investigated
to test the validity of SCHC.

\section{Experimental set-up}
\label{sec-exp}

In this analysis, $J/\psi$ mesons were identified 
with the ZEUS detector at HERA 
by their decays to $ \mu^+\mu^-$ or to $e^+ e^-$.
The muon sample corresponds to an integrated luminosity of $38.0\pm0.6 \pbi$,
collected in 1996 and 1997
when $27.5 \gev$ positrons\footnote{Hereafter, ``positron'' 
is used to refer to both electron and positron beams. 
At the values of $Q^2$ studied here,  
$e^-p$ and $e^+p$ scattering were assumed to give identical results since 
contributions from $Z^{\rm 0}$ exchange are negligible.
Similarly, ``electron'' is used to refer to either the electron or 
positron from the decay of the $J/\psi$.}
were collided with $820 \gev$ protons.
For the measurement with $20<W<30 \gev$, a sample from
an integrated luminosity of $27.5 \pm 0.4 \pbi$ was used.
The electron sample corresponds to an integrated luminosity of 
$55.2\pm 1.2 \pbi$,
collected in 1999 and 2000 in collisions of $27.5 \gev$ positrons
with $920 \gev$ protons.

The ZEUS detector is described in detail elsewhere~\cite{zeusdetector};
only the components most relevant for this analysis are outlined here. 

Charged particles are tracked by the central tracking detector 
(CTD)~\citeCTD,
which operates in a magnetic field of 1.43 T provided by a 
thin super-conducting coil.
The CTD consists of 72 cylindrical drift chamber layers, 
organised in 9 superlayers covering the polar-angle\footnote{
 The ZEUS coordinate system is a right-handed Cartesian system,
 with the $Z$-axis pointing in the proton beam direction,
 referred to as the ``forward direction'',
 and the $X$-axis pointing left towards the centre of HERA.
 The coordinate origin is at the nominal interaction point.
 The pseudorapidity is defined as $\eta=-\ln(\tan \frac{\theta}{2})$,
 where the polar angle, $\theta$, 
 is measured with respect to the proton beam direction.}
region $15^{\circ}<\theta<164^{\circ}$.
The relative transverse-momentum resolution for full-length tracks is 
$\sigma(p_T)/p_T=0.0058p_T\oplus 0.0065 \oplus 0.0014/p_T$,
with $p_T$ in GeV.

Charged particles in the forward direction are detected in
the forward tracking detector (FTD)~\cite{zeusdetector}, which
consists of three planar drift chambers perpendicular to the beam
covering the polar angles $7^{\circ}<\theta<28^{\circ}$.
Each chamber is made of three layers of drift cells; the wire
directions in the second layer are rotated by $120^\circ$ with respect to the 
first layer and similarly for the third layer with respect to the second.
Each drift cell has six sense wires.
Thus each chamber measures a track segment in three projections with up to six 
hits per projection.

Surrounding the solenoid is the high-resolution uranium-scintillator 
calorimeter (CAL)~\citeCAL.
It consists of three parts: 
the forward (FCAL, $2.6^\circ<\theta<36.7^\circ$), 
the barrel (BCAL, $36.7^\circ<\theta<129.1^\circ$)
and the rear (RCAL, $129.1^\circ<\theta<176.2^\circ$) calorimeters.  
Each part is subdivided transversely into towers and longitudinally into one 
electromagnetic section (EMC)
and either one (in RCAL) or two (in BCAL and FCAL) hadronic sections (HAC).
The smallest subdivision of the calorimeter is called a cell. 
Under test beam conditions, the CAL has relative energy resolutions of 
$\sigma(E)/E=0.18/\sqrt{E}$ 
for electrons hitting the center of a calorimeter cell and
$\sigma(E)/E=0.35/\sqrt{E}$ for single hadrons ($E$ in $\Gev$).
Cell clusters were used to aid in the identification of muons and electrons.

The forward plug calorimeter (FPC)~\cite{FPC} is a lead-scintillator 
sandwich calorimeter with wavelength-shifter fibre readout.
Installed in 1998 in the $20\times 20$ cm$^2$ beamhole of the FCAL,
it has a small hole of radius 3.15 cm in the centre 
to accommodate the beampipe.
It extends the pseudorapidity coverage of the forward calorimeter
from $\eta < 4.0$ to  $\eta < 5.0$. 

The small-angle rear tracking detector (SRTD)~\cite{nim:a401:63}
is attached to the front face of the RCAL.
The SRTD consists of two planes of scintillator strips read out
via optical fibres and photomultiplier tubes. 
It covers the region $68 \times 68$ cm$^2$
in $X$ and $Y$ with the exclusion of a $8 \times 20$ cm$^2$ hole at the
centre for the beampipe. 
The SRTD provides a transverse position resolution of 3 mm and was
used to measure the positions of electrons, from the $J/\psi$ decay,
produced at small angles to the positron beam 
direction.

The muon system consists of limited streamer tubes (forward, barrel and
rear muon chambers: FMUON~\cite{zeusdetector}, B/RMUON~\cite{BRMU}) 
placed inside and outside the magnet yoke.  
The inner chambers, F/B/RMUI, cover the polar angles 
between $10^{\circ}<\theta<35^{\circ}$, $34^{\circ}<\theta<135^{\circ}$
and $135^{\circ}<\theta<171^{\circ}$, respectively.
The FMUON has additional drift chambers and 
permits high-momentum muon reconstruction for polar angles
between $6^\circ$ and $30^\circ$ using the magnetic field of 1.7 T
produced by two iron toroids placed at $Z=9$ m and the toroidal field
of 1.6 T provided by the yoke coils. The relative momentum resolution of
$\sigma(p)/p=0.2$, up to $20 \gev$, is dominated by the 
multiple scattering.

The proton-remnant tagger (PRT1)~\cite{prt}
consists of two layers of scintillation 
counters located at $Z=5.15$~m,
and covers the pseudorapidity range \mbox{$4.3<\eta<5.8$}. 
It was used, up to the end of the 1997 running period, to tag events
in which the proton diffractively dissociates.  

The luminosity was determined from the rate of the bremsstrahlung 
process $e\,p\, \rightarrow\, e\, \gamma\, p$, where the photon was
measured with  a lead-scintillator calorimeter~\cite{app:b32:2025} 
located at $Z=-107$ m.

\section{Kinematic variables and reconstruction \label{seckine}}

The kinematic variables used to describe exclusive $J/\psi$ production,
\[
e(k)\,p(P)\, \rightarrow\, e(k')\, J/\psi(v)\, p(P') ,
\]
where 
$k$, $k'$, $P$, $P'$ and $v$ are the four-momenta of the incident
positron, scattered positron, incident proton, scattered proton and
$J/\psi$ meson, respectively, are defined as follows:

\begin{itemize}
\item $Q^2 = -q^2=-(k-k')^2$, the negative four-momentum squared of the
  exchanged photon;
\item $W^2 = (q+P)^2 $, the squared centre-of-mass energy of the
  photon-proton system;
\item $y = (P \cdot q)/(P \cdot k)$, the fraction of the positron
  energy transferred to the photon in the rest frame of the initial-state
  proton;
\item $t = (P - P')^2=(q-v)^2 $, the square of the four-momentum 
  transfer at the proton vertex.
\end{itemize}
The following angles are used to describe the decay of the $J/\psi$:
\begin{itemize}
\item $\Thel$ and $\fhel$, the polar and azimuthal angles of the 
  positively charged decay lepton 
  in the helicity frame, defined as the $J/\psi$ rest frame with 
  the quantisation axis taken to be the $J/\psi$ direction in the 
  photon-proton centre-of-mass system. The origin of the azimuthal angle,
  $\fhel$, corresponds to the case when the decay particles are 
  produced in the production plane, defined as the plane containing 
  the incident photon (assumed to be in the positron direction)
  and the $J/\psi$ momentum vectors.
\end{itemize}
In addition to the above quantities, 
$M_Y$, the mass of the diffractively produced state $Y$, characterises
the major background process, the proton-dissociative reaction 
$e\, p\, \rightarrow\, e\, J/\psi\, Y$.

For the selected events, $Q^2$ ranged from the 
kinematic minimum, $Q^2_{min} = M^2_e y^2/(1-y) \approx 10^{-12} \ \Gev^2$, 
where $M_e$ is the positron mass, 
up to $Q^2_{max} \approx 1 \ \Gev^2$,
the value at which the scattered positron starts to be observed in the CAL,
with a median $Q^2$ of approximately $5 \cdot 10^{-5} \gev^2$. 
Since the typical $Q^2$ is small, it can be neglected in the reconstruction 
of the other kinematic variables.

The photon-proton centre-of-mass energy, $W$, can be expressed as 
$W^2 \approx 2E_p (E - p_{Z})_{ll}$,
where $E_p$ is the laboratory energy of the incoming proton 
and $(E - p_{Z})_{ll}$  is the difference between the energy 
and the longitudinal momentum of the dilepton system.

The squared four-momentum transfer  at the proton vertex is given by
$t \approx -p^2_{T}$, the transverse-momentum squared of the dilepton system. 
Non-zero values of $Q^2$ give $t$ values that differ from $-p^2_T$ 
by less than $Q^2$;
this effect is corrected for using the Monte Carlo simulation.

Since neither the scattered positron nor the scattered proton was
observed, 
the kinematic variables were reconstructed using only the 
measured momenta of the decay particles.
At low $W$, the $J/\psi$ mesons are produced in the forward direction, while
at high $W$ they are produced in the backward direction.
For the $J/\psi\, \rightarrow\, \mu^+\mu^-$ sample, 
the CTD, FTD and FMUON information were used when available.
For the $J/\psi\, \rightarrow\, e^+e^-$  sample, the CTD or the CAL/SRTD 
information was used. 
The relevant {\it in situ} electron-energy resolution of the CAL 
for energies in the range $3<E<20 \gev$ averages 
$\sigma(E)/E=0.27/\sqrt{E}$ for this analysis~\cite{mellado}.

\section{Trigger and event selection \label{sec:sel}}
The events were selected online via a three-level trigger 
system~\cite{zeusdetector}.
The signature for exclusive $J/\psi$ photoproduction events
consists of a pair of charged leptons, 
with no other significant activity in either the CTD or the CAL,
since the scattered positron and proton escape undetected
down the beampipe at small scattering angles. 
For the two decay channels, 
the elasticity cuts described below were imposed.
They restrict
the photon virtuality to $Q^2 \lesssim 1 \gev^2$ 
and the mass of the dissociative system to $M_Y \lesssim 3.0 \gev$.
No cut on $t$ was applied on either channel.
To select candidate events for the muon and electron decay channels,
different selection cuts were applied.
\subsection{Muon channel \label{sec:musel}}
The trigger selected events with at least two tracks in the CTD 
or one track in the CTD and one track in FMUON.
At least one track had to point towards an energy deposit compatible 
with a minimum ionising particle (m.i.p.)
in the CAL and either a hit in the FMUI 
or a segment in the B/RMUI. 
The trigger efficiency for events that passed the offline selection cuts, 
defined below, was determined from independent triggers and from 
MC simulations to be between $50\%$ and $75\%$, depending on $W$.

Events having the following characteristics were selected offline:
\begin{itemize}
    \item exactly two oppositely charged tracks from a common vertex, 
          with $Z$-coordinate $|Z_{vertex}|<50$ cm,
          at least one of which matches either a hit in the FMUI
          or a segment in the B/RMUI;
    \item each CTD track passes through at least three superlayers, 
          effectively limiting the polar-angle region of these ``CTD tracks''
          to $ 17^{\circ}\ \leq\ \theta\ \leq\ 163^{\circ}$;
    \item  the angle between the two tracks is less than 174.2$^{\circ}$, 
          in order to reject cosmic-ray events;
    \item CAL energy associated with each track 
        consistent with the energy deposit of a m.i.p., 
        i.e. between 0.8 and $5 \gev$, 
          with a ratio of at least 0.8 between the energies
          in the HAC and the EMC sections.
          The energy was associated with the track if it was inside a cone
          of radius 30 cm (in EMC) or 50 cm (in HAC), centred at the
          impact position of the track extrapolated on to the CAL;
    \item no CAL cell, apart from those associated with a candidate muon,
          with energy above the threshold level of 150 MeV to 
          200 MeV, depending on the calorimeter part and section.
          This elasticity requirement rejects proton-dissociative 
          and inelastic events as well as 
          deep inelastic scattering (DIS) events. 
          The PRT1 was not used to veto the events.
\end{itemize}

To define the kinematic region in which both decay muons could be
well reconstructed, the analysis was limited to the range $20<\Wgp<170 \gev$.
The requirement that CTD tracks traverse at least three superlayers 
leads to a small acceptance for two such tracks in the region $20<W<30 \gev$.
In this region, therefore, additional events were accepted
with one CTD track and a second track in the FMUON spectrometer, 
which both measures the momentum and triggers the event.
The FMUON track was re-fitted to the vertex including the FTD segments 
to improve the reconstruction parameters.
In the kinematic region $30<W<170 \gev$, only events with two 
CTD tracks were used. 

\subsection{Electron channel}
The $J/\psi\, \rightarrow\, e^+ e^-$ events were selected online 
using two different trigger algorithms:
\begin{itemize}
\item
the first algorithm was optimised for events with one or two tracks
in the CTD. It required at least one track, but fewer than five tracks, 
in the CTD, 
a total energy deposit greater than $1.8 \gev$ in an EMC section of the CAL
and an energy
of less than 3.75 GeV in the region of $60 \times 60$ cm$^2$ of FCAL around 
the forward beam pipe;
\item
the second algorithm was optimised for events with zero tracks in the CTD.
It required at least two deposits in the RCAL EMC,
each with an energy greater than $2.1 \gev$. 
\end{itemize}
The trigger efficiency for events that passed the offline selection cuts, 
defined below, was determined from independent triggers
and from MC simulations to be between $80\%$ and $90\%$, depending on $W$.

The following offline selection requirements were applied:
\begin{itemize}
      \item for events with a reconstructed tracking vertex,
        the cut $|Z_{VTX}|<50$ cm was applied.
        Events without a vertex were accepted and were assigned
        a vertex $Z$ position corresponding to the nominal interaction point;
    \item the CAL energy deposits were grouped into clusters. Events were
          selected for further analysis if they satisfied one of the following
          criteria:
       \begin{itemize}          
       \item for events with two CTD tracks,
          the CAL energy associated with each track, 
          inside a cone of radius 25 cm
          centred at the impact position of the track extrapolated on 
          to the CAL,
          had to be consistent with
          the energy deposition expected for an electron:
          $E_{EMC} / E > 0.9 $, where $E$ and $E_{EMC}$ 
          are, respectively, the total energy 
          and the energy deposited in the EMC section.
          These two tracks were then considered electron candidates;
        \item for events with one CTD track and one cluster not associated
          with it, the CAL energy associated with
          the track had to have $E_{EMC} / E > 0.9 $ and
          the CAL energy of the unassociated cluster
          had to be more than $3\gev$, with
          $E_{EMC} / E > 0.98 $, to be considered an electron candidate;   
        \item for events with no CTD track, 
          two clusters, each with CAL energy more than $3.5\gev$ 
          and with $E_{EMC} / E > 0.98 $, were required as candidate electrons;
\end{itemize}
    \item any energy deposit in the CAL cells, 
          not associated with either of the two electron candidates, 
          was required to be less than either 
          200 MeV or 300 MeV, depending on the calorimeter part and section.
          This requirement\footnote{This elasticity requirement 
          is less stringent than the one used 
          for the selection of the muon sample because of an increased 
          noise level in the CAL during the data taking in 1999 and 2000.}
rejects proton-dissociative and inelastic events, as well as DIS events;
    \item further to reduce events from proton dissociation,  
          the energy measured in the FPC was required to be less than $1 \gev$.
\end{itemize}

The analysis was restricted to the kinematic region $20<\Wgp<290 \gev$ because
the acceptance drops at lower $\Wgp$ and the QED-Compton background dominates
at higher $\Wgp$.

%
\section{Monte Carlo simulation\label{sec:acceptance}} 
The acceptance and the effects of the detector response were determined using
samples of Monte Carlo (MC) events.
The ZEUS detector response to the generated particles
was simulated in detail using a program based on GEANT3.13\cite{geant}.
All the generated events were processed through the same reconstruction 
and analysis chain as the data.

The exclusive processes $e \, p \, \rightarrow e \, J/\psi \,p$ and
$e \, p \, \rightarrow e \, \psi(2S) \,p$ 
were modelled using the MC generators DIPSI~\cite{DIPSI,zfp:c57:89,*ryskin2}
and ZEUSVM~\cite{thesis:muchorowski:1996}.
The events were weighted with a $\gamma p$ 
cross section proportional to $W^{\delta}$
and with an exponential $t$ dependence $e^{bt}$.
The weight parameters, $\delta=0.70$ and $b=4.3\gev^{-2}$,
were chosen so as to
describe the $\Wgp$ and $t$ dependence of the data,
as discussed in Sections~\ref{sec:cross-section} and~\ref{sec:traj}.
The decay of the $J/\psi$ mesons in the centre-of-mass system was generated 
with a $(1+\cos^2\theta_h)$ distribution, consistent
with the measurement presented in Section~\ref{sec:angle}.
The effect of the initial-state radiation on the acceptance,
estimated using HERACLES 4.6.1~\cite{cpc:69:155,thesis:abe:2001},
was $5\%$ in the region $20<\Wgp<30\gev$ and negligible at larger $\Wgp$.
The final-state radiation of hard photons at large angles with respect 
to either of the two decay leptons decreases the acceptance,  
as these events are rejected by the elasticity requirements applied offline.
The effect was estimated using PHOTOS~\cite{cpc:79:291,thesis:abe:2001}
and was found to be $1.3\%$ in the muon analysis and up to $8\%$
in the electron analysis.
The acceptances, $\cal A$, corrected for these effects,
are given in Table~\ref{tab:xsection}.
They were calculated as the number of events reconstructed in a bin
divided by the number of events generated in the same bin.

Proton-dissociative events, $e\,p\, \rightarrow\, e\, J/\psi\, Y$,
were modelled using the generator EPSOFT ~\cite{Kasprzak,*adamczyk},
which simulates $\gamma p$ interactions assuming the exchange of 
the soft-Pomeron trajectory.
At fixed $\Wgp$ and $t$, it models
the mass spectrum of the baryon system $Y$ according to
$d\sigma/dM^2_Y \propto M^{-\beta}_Y$, with $M_Y>1.25 \gev$.
The multiplicity distribution of hadrons from the decay of the 
proton-dissociative system and 
their transverse momenta with respect to the proton-Pomeron collisions axis
are simulated to describe the ZEUS photoproduction data
and hadron-hadron single-diffractive results. 
The longitudinal momenta are generated using a uniform distribution
in rapidity.
The simulation parameters $\delta=0.70$, $b=0.65\gev^{-2}$ and $\beta=2.6$
were chosen to describe the $\Wgp$, $t$ 
and $M_Y$ dependence of the data, as discussed in Section~\ref{sec:p-diss}.
For the region $M_Y < 2 \gev$, $b = 4.0\gev^{-2}$ was used. This
reflects the steeper $t$ distribution observed in low-mass 
hadron-hadron diffraction~\cite{hadhadt,np:b108:1}.  

The  QED $\gamma \gamma\, \rightarrow\, l^+l^-$ background was simulated using
the LPAIR~\cite{Vermaseren} generator.
The QED-Compton scattering, $e\,p\, \rightarrow\, e\, \gamma\, p$, background 
was
simulated using the COMPTON2~\cite{Compton} generator.

\section{Mass spectra and background subtraction \label{background}}
The invariant-mass spectra for the muon- and electron-pair candidates, 
after all offline cuts, 
are shown in Figs.~\ref{fig:mass_spectra_mm} and~\ref{fig:mass_spectra_ee} 
for representative $\Wgp$ bins.
The mass resolution is excellent in the region
$30<W<150 \gev$, where both reconstructed leptons pass through all CTD
layers, and decreases at lower and higher W values, where the leptons are
produced in the forward and rear direction, respectively, at the edges
of the CTD acceptance. In  the analysis of the electron decay channel,
the kinematic region was extended  at very high $W$ by reconstructing the
electrons in the RCAL and SRTD. The mass spectra of the electron pairs have
a tail at low mass due to photon bremsstrahlung.

The final samples contain backgrounds from non-resonant sources: QED
$\gamma \gamma$ processes, QED-Compton scattering and other non-resonant
background (dominated by misidentified pion production),
as well as from resonant processes: 
diffractive $\psi(2S)$ production and proton dissociation.
Contamination from non-diffractive $J/\psi$ production,
$e\,p\, \rightarrow\, e\, J/\psi\, X$, estimated using
the MC generator HERWIG 5.8~\cite{cpc:67:465}, is negligible.

\subsection{Non-resonant background \label{sec:non-res}}
\begin{figure}[htbp!]
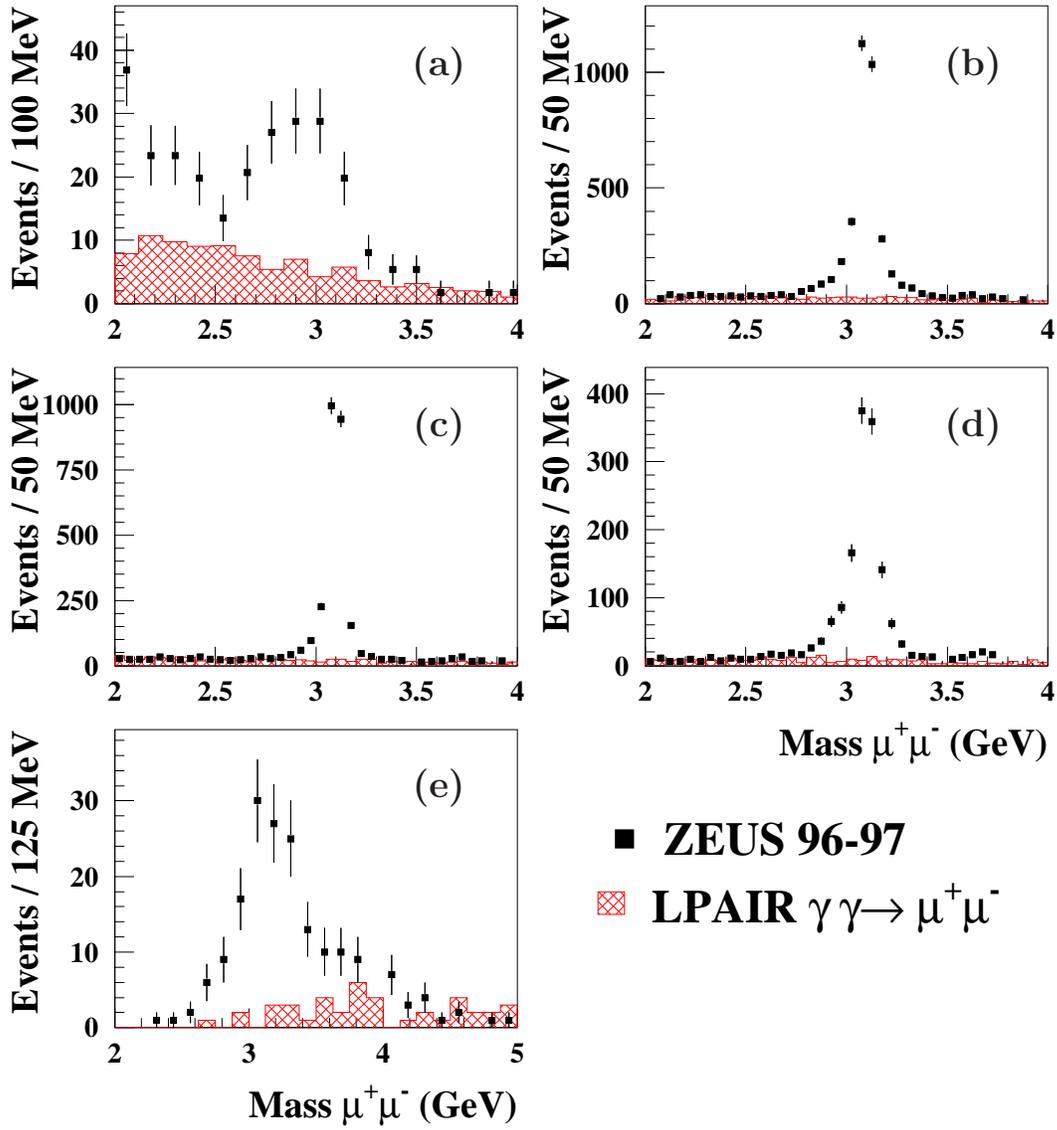

\psfigadd{fig.1}{17cm}{17cm}{
\Text(720,1450)[]{\large\bf (a)}
\Text(1430,1450)[]{\large\bf (b)}
\Text(720,970)[]{\large\bf (c)}
\Text(1430,970)[]{\large\bf (d)}
\Text(720,490)[]{\large\bf (e)}
}
\caption{
Invariant-mass distributions of the $\mu^+\mu^-$ 
pairs in the different $\Wgp$ regions:
(a) $20<W<30 \gev$, (b) $30<W<70 \gev$, (c) $70<W<110 \gev$, 
(d) $110<W<150 \gev$ and (e) $150<W<170 \gev$.
The histograms represent the LPAIR 
distributions of the non-resonant background.
}
\label{fig:mass_spectra_mm}
\end{figure}

\begin{figure}[htbp!]
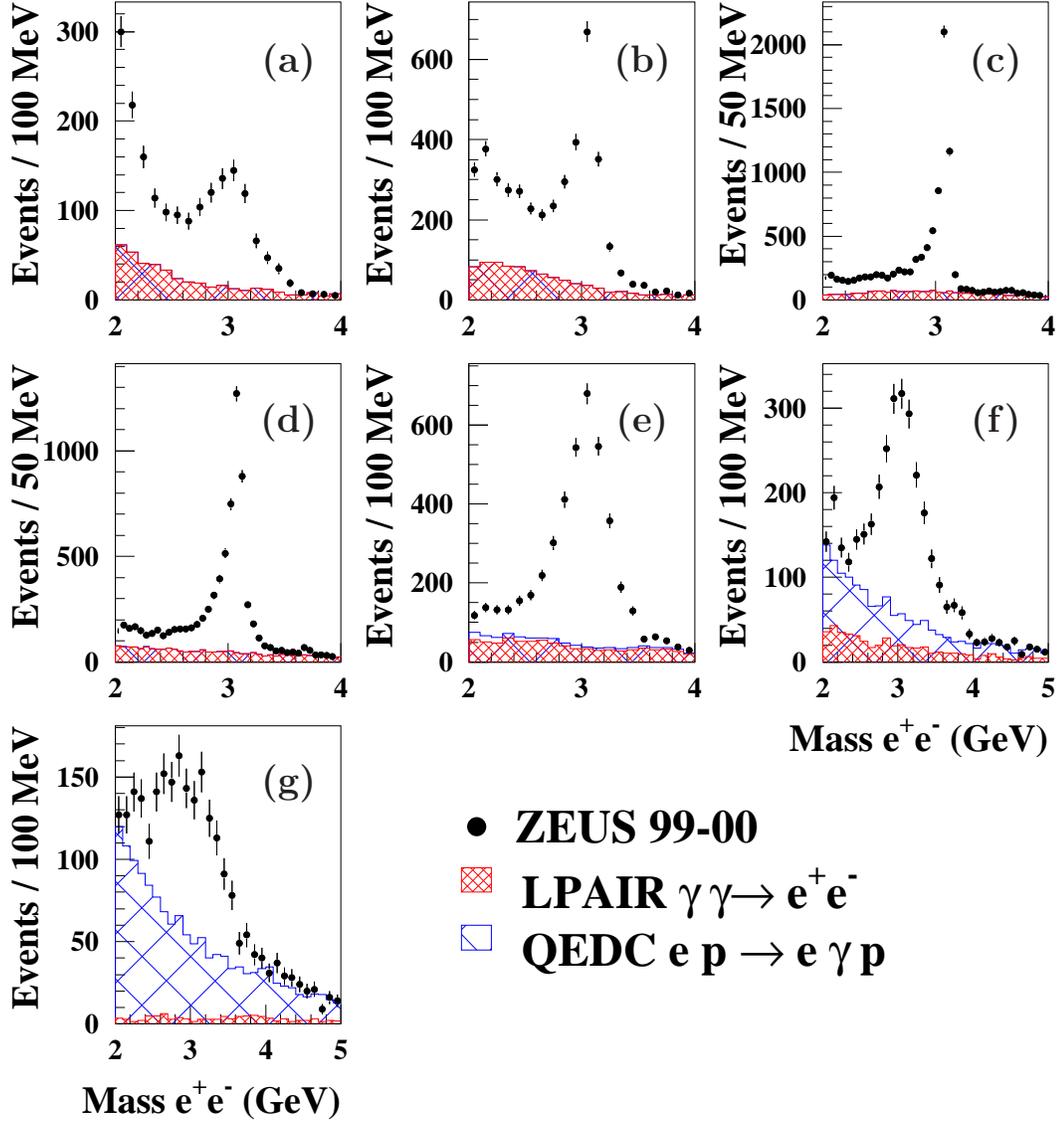

\psfigadd{fig.2}{17cm}{17cm}{
\Text(520,1450)[]{\large\bf (a)}
\Text(990,1450)[]{\large\bf (b)}
\Text(1460,1450)[]{\large\bf (c)}
\Text(520,970)[]{\large\bf (d)}
\Text(990,970)[]{\large\bf (e)}
\Text(1460,970)[]{\large\bf (f)}
\Text(520,490)[]{\large\bf (g)}
}
\caption{
Invariant-mass distributions of the 
$e^+e^-$ pairs in the different $\Wgp$ regions:
(a) $20<W<35 \gev$, (b) $35<W<50 \gev$, (c) $50<W<90 \gev$, (d) $90<W<140 
\gev$, (e) $140<W<200 \gev$, (f) $200<W<260 \gev$ and (g) $260<W<290 \gev$. 
The close-hatched histogram represents the LPAIR Monte Carlo
distribution for the non-resonant background
and the wide-hatched histogram that from COMPTON2.
}
\label{fig:mass_spectra_ee}
\end{figure}

Background from the QED process $\gamma \gamma\, \rightarrow\, l^+l^-$,
in which a lepton pair is produced by the fusion 
of a photon radiated by the positron with a photon radiated by the proton,
was estimated 
using the LPAIR MC normalised to the data in the mass region 
outside the $J/\psi$ and $\psi(2S)$ resonances.
It is shown in the dilepton-mass spectra of 
Figs.~\ref{fig:mass_spectra_mm} and~\ref{fig:mass_spectra_ee}.
The typical contribution in the signal region
is $10\%$ for the muon sample
and up to $20\%$ for the electron sample.

Pions produced at low angles in the forward direction 
can be misidentified as muons or electrons.
This background dominates at low masses and for $W < 50 \gev$, 
as can be seen in Figs.~\ref{fig:mass_spectra_mm}(a) 
and ~\ref{fig:mass_spectra_ee}(a,b).

For the electron sample, there is an additional contribution from
QED-Compton scattering with initial-state radiation.
It was estimated using the COMPTON2 MC normalised to the data 
in the mass region outside the resonances for $W>230 \gev$, where it
dominates the background distribution, as shown 
in Fig.~\ref{fig:mass_spectra_ee}.
The size of this background ranges from $3\%$ at $W= 200 \gev$ 
to $50\%$ at $W = 275 \gev$. 
The total non-resonant contributions are given in Table \ref{tab:xsection}.

\subsection{Events from $\psi(2S)$ production}
Events from $\psi(2S)$ production can fake exclusive $J/\psi$ events, 
mainly through the decay
$\psi(2S)\,\rightarrow\, J/\psi\, +\, neutrals$, 
with a branching ratio ${\cal B}=(23.1\pm2.3)\%$~\cite{PDG00},
when the $J/\psi$ decays into two leptons and the neutral particles
are not detected in the CAL.
An additional source of background comes directly from
$\psi(2S)\,\rightarrow\, l^+l^-$ decays,
with a branching ratio ${\cal B}=(1.03\pm0.35)\% $~\cite{PDG00},
since the mass window used to count the $J/\psi$ signal, for $\Wgp \lap$ 35
GeV and $\Wgp \gap$ 140 GeV,
is large enough to include the $\psi(2S)$ mass at $3.685 \gev$.
The number of events from $\psi(2S)$ production present in the $J/\psi$ 
elastic sample was estimated using the $\psi(2S)$ DIPSI MC events and the
ratio of production cross sections, 
$\psi(2S)/ (J/\psi)=0.150\pm0.027(stat.)\pm0.022(syst.)$~\cite{psi2s},
to be smaller than $7\%$, 
as shown in Table~\ref{tab:xsection}.

\subsection{Proton dissociation\label{sec:p-diss}}
The largest source of background is given by the
diffractive production of $J/\psi$ mesons with proton dissociation,  
$e\,p\, \rightarrow\, e\, J/\psi\, Y$,
when the system $Y$ has a small mass and its decay
products are not detected in either the FCAL, the PRT1 or the FPC. 

To estimate this background, the elasticity cut was
removed in the region of fragmentation of the system $Y$; the 
proton-dissociative data obtained in this way were used to tune the
EPSOFT MC generator. The data and the fraction of MC events in which
energy was deposited in the PRT1 or the FPC were then used
to estimate the number of proton-dissociative events contaminating
the exclusive $J/\psi$ sample. 

The parameters $b$ and $\beta$ of the EPSOFT MC generator 
were tuned using a sample of dimuon events triggered by the B/RMUI. 
These data were selected as described in Section \ref{sec:musel}
but, for those FCAL cells with $\theta<30^\circ$, 
the elasticity requirement was removed
and an energy of at least 300 MeV in both EMC and HAC sections was required.
A rapidity gap of at least $\Delta \eta>1.3$ between the $J/\psi$ meson 
and the products of the proton-dissociative system was required.
The final sample of $J/\psi$ candidates,
in the kinematic region $90<W<130\gev$, consisted of about 600 events.
The data sample corresponds, according to MC simulations, to
the region $3.5<M_Y<30\gev$ and $p^2_T<10\gev^2$.
The $b$ slope and $\beta$ parameters were determined to be 
$0.65\pm0.10 \gev^{-2}$ and $2.6\pm0.3$, respectively, 
from the study of the $p^2_T$ and visible CAL-energy distributions.
The $\Wgp$ distribution for the proton-dissociative 
process was consistent with that of the exclusive channel.
A similar study performed for the electron-decay channel, 
using a smaller sample of data triggered without the veto on energy deposited 
in the FCAL region around the beam pipe,
yielded values of $b$  and $\beta$ 
consistent with those determined from the muon sample.

The proton-dissociative events misidentified as exclusive $J/\psi$ production
were subtracted in each ($\Wgp$, $t$) bin 
for the cross sections presented here, 
for both the muon and the electron analyses.

In the muon analysis, the contribution from proton-dissociative events 
in the elastic sample was estimated using the PRT1. In each $W$ and $t$ bin,
the quantity $f_{p-diss} = f^{data}_{PRT1} \cdot \frac {1} {\epsilon}$,
was computed,
where $f^{data}_{PRT1}$ is the fraction of the data tagged in the PRT1
and $\epsilon$ is the MC tagging efficiency, 
defined as the probability to obtain a tag in PRT1 in EPSOFT events. 
Using $f^{data}_{PRT1}$ = 12.6\% and $\epsilon=57.2\%$, 
$f_{p-diss}$ was estimated to be $(22.0\pm 2.0(stat.) \pm 2.0(syst.))\%$ 
for $|t|<2\gev^2$, with no $W$ dependence.
The fraction of $f_{p-diss}$ was estimated to increase with $t$ from
$(11.0^{+3.1}_{-1.4}(stat.+syst.))\%$, in the first $t$ bin,
up to $(49^{+14}_{-6}(stat.+syst.))\%$, for $1.2<|t|<1.8\gev^2$.

In 1998, the PRT1 was no longer used. The FPC was inserted 
and was used to veto proton-dissociative events.
In the electron analysis, which uses the data taken in 1999 and 2000,
the amount of proton-dissociative background in the elastic sample
was estimated from
$f_{p-diss} = f^{data}_{FPC} \cdot (\frac {1}{\epsilon} - 1)$ and
found to be $(17.5\pm 1.3(stat.) ^{+3.7}_{-3.2}(syst.))\%$, 
using $f^{data}_{FPC}=12\%$ for the FPC-tagged events 
and $\epsilon=40.7\%$,
where $\epsilon$ is defined as the probability to obtain a tag in FPC
in EPSOFT events. The formula is different with respect to the muon case
because the events tagged by the FPC were also rejected.
The fraction of $f_{p-diss}$ was estimated to increase with $t$ from
$(6^{+3}_{-4}(stat.+syst.))\%$ at low $t$ 
to $(28^{+8}_{-5}(stat.+syst.))\%$ at $0.85<t<1.15 \gev^2$.

As an independent check on the estimation of the proton-dissociative 
background, an alternative model to EPSOFT was used. 
In the baryon resonance region, at low $M_Y$, a resonant component with 
slope $b=6.5$ GeV$^{-2}$ was considered. A second component due to 
non-resonant proton dissociation with slope $b=0.65$ GeV$^{-2}$ was added.
The two  components  were constrained  to satisfy
the first moment of the  finite-mass sum rule~\cite{pr:74:1}.
This model yielded results that agreed with those from EPSOFT to
within $2\%$ for both the cross section and the $b$ slope.

\subsection{Signal determination \label{sec:masssignals}}

Since the mass spectra have shapes and background contributions that vary
with $\Wgp$, $t$ and the decay channel, different procedures were used in
the muon and electron analyses to determine the number of signal events.

For the muon analysis, the signal events were counted in a mass window 
corresponding to $\pm$ three standard deviations of the Gaussian fit
from the mean fitted value of the $J/\psi$ mass. Since the mass resolution 
depends on the kinematic region, the mass windows were different 
for different $\Wgp$ and $t$ bins. 
The number of $\gamma \gamma\, \rightarrow\, \mu^+\mu^-$ 
background events was 
estimated as described in Section~\ref{sec:non-res} and
subtracted in each mass window to obtain the number of $J/\psi$ candidate 
events.
For the lowest W bin, there is a remaining background,
coming from misidentified pions.
The mass spectrum was then fitted to the sum of the distribution 
predicted by the signal MC and a single exponential function for the 
background.

For the electron analysis, the backgrounds due to QED-Compton scattering
and from $\gamma \gamma\, \rightarrow\, e^+e^-$ were subtracted. 
The remaining mass spectrum was then fitted, in each bin,
to the sum of the distribution predicted by the signal MC
and a single exponential function for the remaining background.
The latter is dominated by misidentified pions.
The MC gives a good description of the M$_{e^+e^-}$ 
distributions observed in the data with typical $\chi^2$/ndf better than 1.5
%
\section{Systematic uncertainties \label{sec:systematics}}
%
The systematic uncertainties on the cross sections are given separately 
for the two decay channels. 
For the muon channel, the following sources of uncertainty were considered.
\begin{itemize}
\item trigger efficiency: 
the uncertainty due to that of the trigger efficiency
was $\pm 5\%$ for the CTD track reconstruction 
and up to $\pm 7\%$ for the muon selection; 
\item event selection:
the minimum number of required CTD superlayers was raised 
from three to four;
the cut on the angle between two tracks was relaxed from 
$174.2^{\circ}$ to $176.4^{\circ}$;
the criteria for the identification of a m.i.p.\ in the CAL were varied
and the cell energy threshold for the selection 
of exclusive events was increased from 150 or 200 MeV to 300 MeV.
The resulting uncertainty  was $\pm 3\%$.
For events in the range $20<W<30 \gev$, 
the uncertainty due to the track reconstruction in the FMUON/FTD was
$\pm 11\%$;
\item MC model dependence: 
the uncertainty was estimated by varying the parameters 
$b$ and $\delta$ of the DIPSI MC simulation 
within the range $4.1<b<4.5 \gev^{-2}$ and $0.60<\delta<0.75$.
The centre-of-mass decay angular distribution of the muons was changed
to $[1+\alpha+(1-3\alpha)\cos^2\theta_h]$ with $\alpha = -0.05$,
consistent with the measurement presented in Section \ref{sec:angle}.
The overall uncertainty due to model dependence was $\pm 5\%$;
\item proton-dissociative subtraction: 
the uncertainty on the modelling of the hadronic final state in 
proton-dissociative events was estimated to be $\pm 2\%$
by varying the parameters of the simulation 
by $\pm 0.3$ for $\beta$ and $\pm0.10\gev^{-2}$ ($\pm 2 \gev^{-2}$)
for $b$ when $M_Y > 2 \gev$ ($M_Y<2 \gev$), 
as discussed in Section \ref{background};
\item non-resonant background subtraction:
the uncertainty was typically $\pm 2\%$.
In the lowest $W$ bin, the uncertainty was $7\%$, as determined by the fit;
;
\item the uncertainty in the luminosity determination was $\pm 1.7\%$  
for the 1996-1997 running period.
\end{itemize}
For the electron channel, the following 
sources of uncertainty were considered:
\begin{itemize}
\item trigger efficiency: 
the estimated uncertainty was $\pm 2.5\%$ for the CTD, 
$\pm (1-5)\%$, depending on $\Wgp$,
 for the CAL energy threshold and $\pm 3\%$ for the trigger stream
requiring two isolated electromagnetic clusters;
\item event selection:
the effect of varying the elasticity requirements by raising
the threshold by 100 MeV was $-2.5\%$ to $+2.5\%$, depending on $W$;
\item MC model dependence: 
the uncertainty, 
estimated in the same way as described for the muon channel, 
was less than $\pm 2.5\%$;
\item proton-dissociative subtraction: 
the uncertainty, estimated as described for the muon channel,
was $^{+4.0}_{-3.5}\%$;
\item non-resonant background subtraction: 
the uncertainty in the normalisation of this background,
as determined by the fit, varied between 1 and $6\%$,  depending on $W$;
\item the uncertainty in the luminosity determination was 
$\pm 2.25\%$ in the years 1999 and 2000.
\end{itemize}
The overall systematic uncertainty was determined 
by adding the uncertainties in quadrature.
An additional uncertainty of $1.7\%$~\cite{PDG00}
associated with the branching ratio ${\cal B}_{J/\psi\, \rightarrow\, l^+l^-}$
was not included. 
Since the major sources of systematic uncertainty
are mostly independent of $W$, they have a small influence on the
determination of $\delta$ and the Pomeron trajectory.
%
\section{Decay angular distributions \label{sec:angle}}
%
Since the decay angular distributions were used to reweight the MC simulated
events and thus affect the cross-section measurements, 
they are discussed first.
They were used to investigate the helicity structure of $J/\psi$ production.
The decay angular distribution is a function of $\Thel$ and $\fhel$, 
the polar and azimuthal angles  of the positively charged lepton 
in the helicity frame.
The normalised angular distributions can be expressed~\cite{wolf}
in the form
\begin{equation}
  \frac{1}{N} \frac{dN}{d \cos \theta_h} = \frac{3}{8} \left[
                               1 + r^{04}_{00} + 
                 \left( 1 - 3 r^{04}_{00} \right) \cos^2 \theta_h
                                       \right]
\label{hel_theta}
\end{equation}
and
\begin{equation}
  \frac{1}{N} \frac{dN}{d \phi_h} = \frac{1}{2 \pi} \left[
                               1 + r^{04}_{1-1} \cos 2 \phi_h 
                                       \right],
\label{hel_phi}
\end{equation}
where the $J/\psi$ spin-density matrix element \rzfzz \ represents the 
probability that the produced $J/\psi$ \ has helicity $0$ and
\rzfpm\ is related to the interference between the non-flip and 
double-flip amplitudes.  
If the $J/\psi$ retains the helicity of the almost-real photon,
as in the hypothesis of $s$-channel helicity conservation (SCHC),
\rzfzz\ and  \rzfpm\ should both be approximately zero.
%

\begin{figure}[htbp!]
\epsfxsize=16cm
\centerline{\epsffile{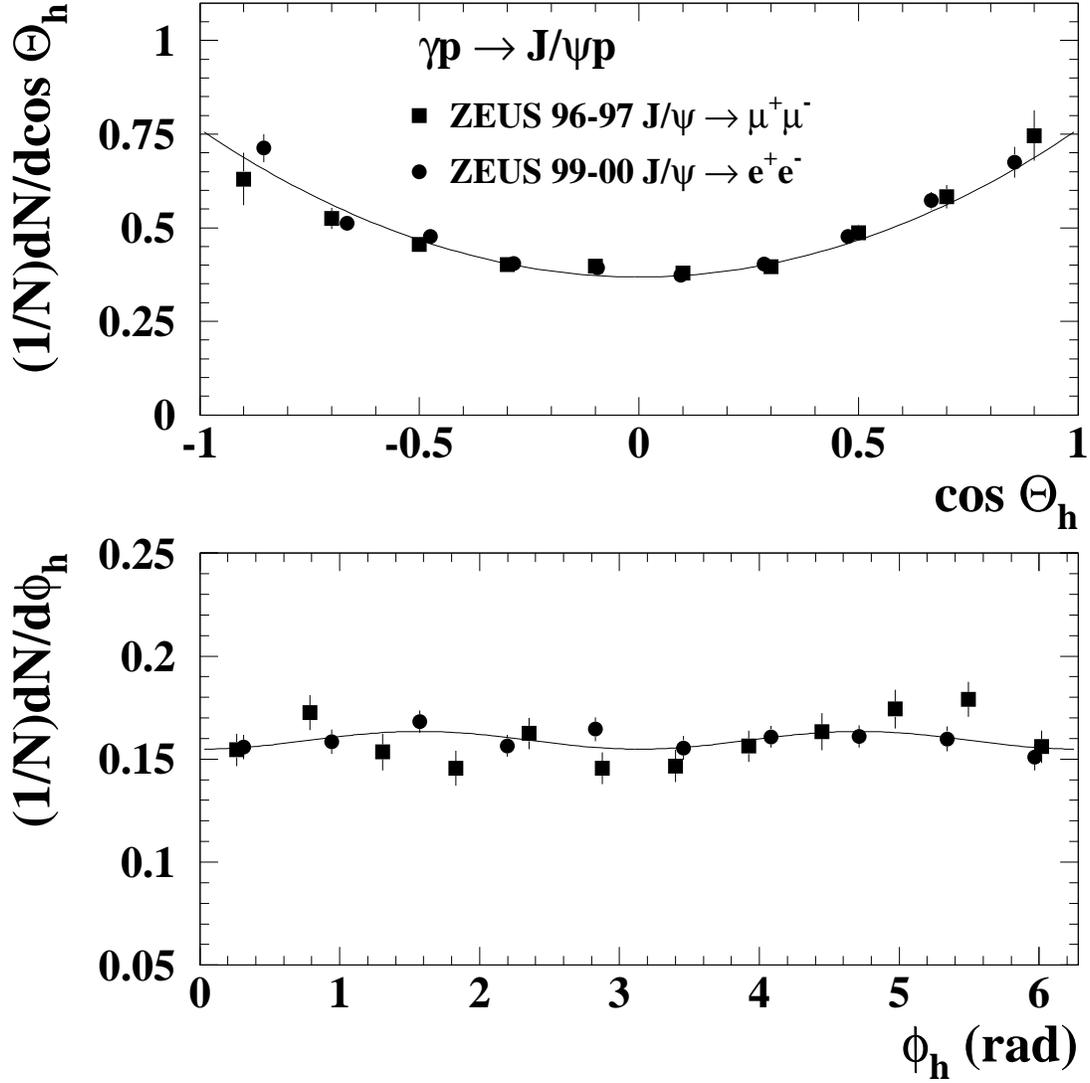}}
\caption{
The acceptance-corrected decay angular distributions for
exclusive $J/\psi$ photoproduction in the kinematic range 
$30<\Wgp<170 \gev$ and $|t|<1 \gev^2$.
The non-resonant background has been subtracted.
The results of both the $\mu^+ \mu^-$ and $e^+e^-$ 
decay channels are presented.
The vertical bars indicate the statistical uncertainties only.
The curves are the results of the fits to Eqs.~(\ref{hel_theta}) 
and (\ref{hel_phi}), as described in the text.
}
\label{fig:helicity}
\end{figure}

The angular distributions of the leptons from $J/\psi$ decay 
are presented in Fig.~\ref{fig:helicity}. They were measured 
in the kinematic range $30<W<170 \gev$ and $|t|<1 \gev^2$,
using events in which both leptons were measured in the CTD.
The non-resonant background was subtracted in each angular bin. 
No subtraction of the dissociative contribution was made, since
the proton-dissociative sample, discussed in Section \ref{sec:p-diss},
displayed similar angular distributions to the elastic events.
The $\psi(2S)$ events were assumed to have the same angular distribution
as the $J/\psi$ events and were not subtracted.

The elements \rzfzz\ and \rzfpm, obtained
by fitting the acceptance-corrected $\theta_h$ and $\phi_h$ distributions
to Eqs. (\ref{hel_theta}) and (\ref{hel_phi}), are
\[
\rzfzz\ = -0.017\pm 0.015(stat.) \pm 0.009(syst.)
\]
and
\[
\rzfpm\ = -0.027\pm 0.013(stat.) \pm 0.005(syst.).
\]
Thus, to within two standard deviations, the SCHC hypothesis holds, as 
expected for heavy mesons \cite{np:b296:569}.
%
\section{$W$ dependence of the cross section \label{sec:cross-section}}
%
The $\gamma p$ cross section for exclusive $J/\psi$ production 
was evaluated from the $ep$ cross section using the expression
\begin{equation}
\sigma_{\gamma\, p\, \rightarrow\, J/\psi\, p} =
\frac{\sigma_{ep\rightarrow e\, J/\psi\, p}}{\Phi} 
= \frac{1}{\Phi} \cdot
\frac{(N_{obs} - N_{non-res}- N_{\psi(2S)} )\cdot( 1 - f_{p-diss})}
     {{\cal L} \cdot {\cal A} \cdot {\cal B}},
\label{eq:sigmaweq}
\end{equation}
where $\Phi$ is the effective photon flux~\cite{flux},
$N_{obs}$ is the number of events in the signal mass region,
$N_{non-res}$ is the number of non-resonant background events,
$N_{\psi(2S)}$ is the number of events from $\psi(2S)$ production, 
$f_{p-diss}$ is the fraction of proton-dissociative events,
${\cal L}$ is the integrated luminosity,
${\cal A}$ is the acceptance 
and
${\cal B}$ is the branching ratio, 
where ${\cal B}=(5.93\pm0.10)\%$ for the electron channel 
and $(5.88\pm0.10)\%$ for the muon channel~\cite{PDG00}.

\begin{figure}[htbp!]
\epsfxsize=16cm
\centerline{\epsffile{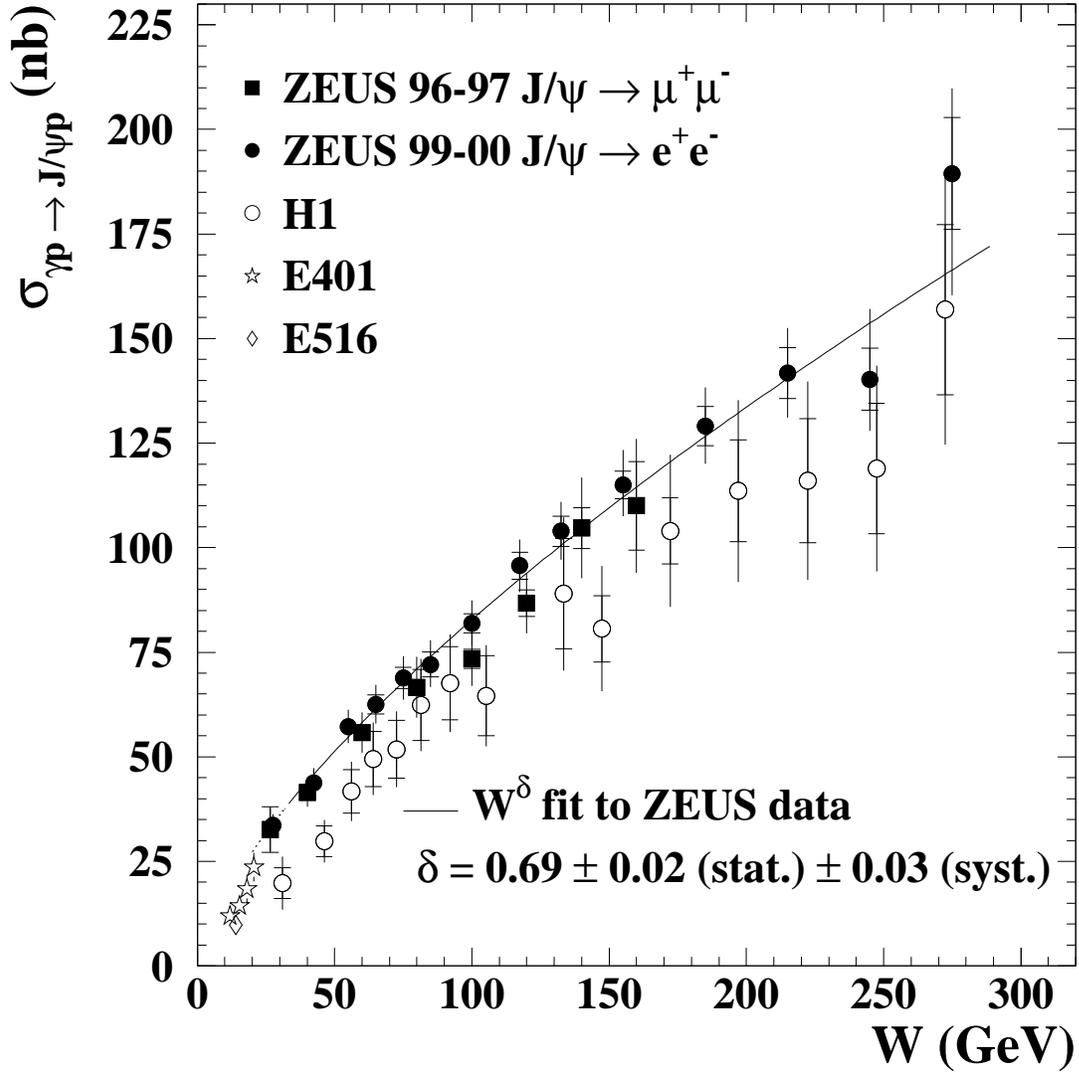}}
\caption{ 
The exclusive $J/\psi$ photoproduction cross section as a function of 
$\Wgp$ for 
$J/\psi \rightarrow \mumu$ and $J/\psi \rightarrow e^+e^-$. 
The inner bars indicate the statistical uncertainties; the outer bars 
are the statistical and systematic uncertainties added in quadrature. 
Results
from the H1~\pcite{H1001}, E401~\pcite{E401} 
and E516~\pcite{E516}  experiments are also shown.
The solid line is the result of a fit to the ZEUS data of the form
$\sigma \propto (W/90 \gev)^\delta$ and the dotted line 
is the extrapolation of the fit.
}
\label{fig:cross-fit}
\end{figure}

The numbers of events, the acceptance, the flux factors 
and the cross sections are given 
in $W$ bins for each decay mode in Table~\ref{tab:xsection} .
The cross section is shown as a function of $\Wgp$ in Fig.~\ref{fig:cross-fit}.
No cut on $t$ was applied.
The small difference ($\sim 4\%$) 
in the normalisation of the muon and the electron 
values is within the correlated uncertainty associated with each decay channel.
The values are larger than those determined previously 
by ZEUS~\cite{ZEUS971}.
The differences are due to a better understanding of the acceptance
and trigger efficiency
and of the background subtraction. Because of these improvements, the results
of this paper supersede those of the previous publication.

Results from the H1 Collaboration~\cite{H1001} and
from fixed-target experiments~\cite{E401,E516} are also displayed in 
Fig.~\ref{fig:cross-fit}.
While the $W$ dependence is similar, there is a normalisation difference
between the H1 and ZEUS values.

The results of fits of the form $\sigma \propto (\Wgp/90 \gev)^{\delta}$ 
to the muon data
and, separately, to the electron data
are given in Table~\ref{tab:results}.
A common fit to the data with $W>30 \gev$, 
including both the muon and the electron measurements, 
with statistical and systematic uncertainties added in quadrature,
yields a value of 
$\delta = 0.69 \pm 0.02 (stat.) \pm 0.03 (syst.)$.  
This result, shown as the curve in Fig.~\ref{fig:cross-fit}, confirms 
the strong energy dependence of the cross section 
observed previously~\cite{ZEUS971,H1001}.
The measurements for $\Wgp<30 \gev$ were not included in this fit
to avoid possible effects due to the charm production 
threshold~\cite{proc:had:1996:170,*fiore}.
However, the result of a fit including the points with $W < 30$ GeV 
does not significantly change the fitted value of $\delta$.

\begin{figure}[htbp!]
\epsfxsize=16cm
\centerline{\epsffile{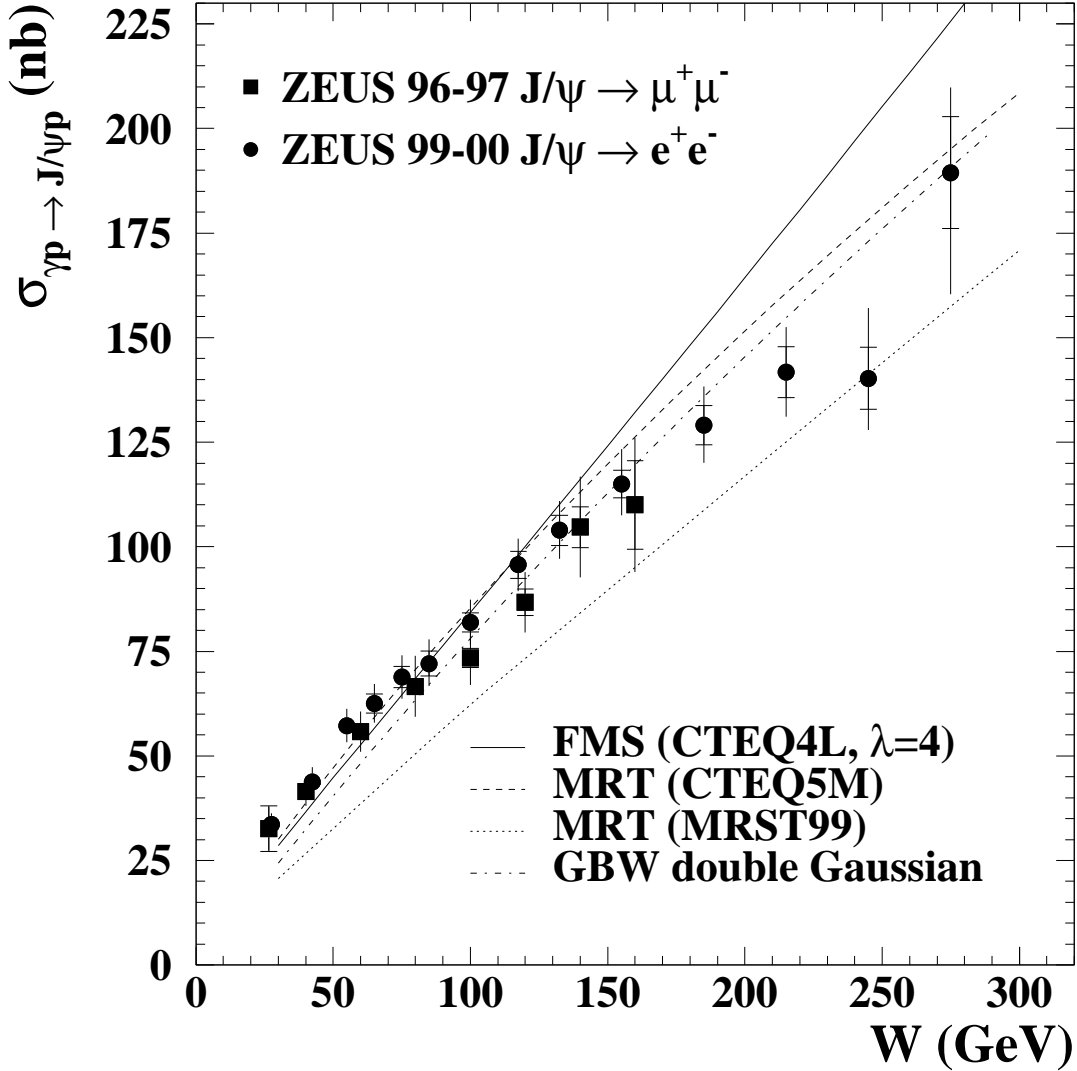}}
\caption{
The exclusive $J/\psi$ photoproduction cross section as a function of 
$\Wgp$ obtained in the two leptonic decay channels,
$J/\psi \rightarrow \mumu$ and $J/\psi \rightarrow e^+e^-$.
The inner  bars indicate the statistical uncertainties, the outer bars 
are the statistical and systematic uncertainties added in quadrature. 
The experimental results are compared to the QCD predictions of 
MRT~\pcite{MRT99}, 
using two different parameterisations of the gluon PDF
in the proton, MRST99~\pcite{mrst99} (dotted curve) and CTEQ5M~\pcite{cteq5m}
(dashed curve). 
The solid curve
shows the QCD prediction of FMS~\pcite{jhep:103:45} using $\lambda = 4$ 
and the CTEQ4L~\pcite{cteq4l} PDF.
The dash-dotted curve displays the prediction~\pcite{soares00} based on the 
colour-dipole model~\pcite{pr:d59:014017,*pr:d60:114023}
with a double-Gaussian $J/\psi$ wave-function;
this prediction was re-scaled to the $b$-slope measured in this paper.
}
\label{fig:cross-mrtcdm}
\end{figure}

The ZEUS data are compared in Fig.~\ref{fig:cross-mrtcdm} to 
leading-log-approximation (LLA) pQCD calculations~\cite{MRT99}, based on 
open $c\bar{c}$  production and parton-hadron duality,
using the CTEQ5M~\cite{cteq5m} (dashed curve) or
MRST99~\cite{mrst99} (dotted curve) parton-density functions (PDF). 
The gluon density is evolved using ``skewed" evolution 
equations~\cite{pl:b462:178,*zfp:c12:263}. 

The solid curve in Fig.~\ref{fig:cross-mrtcdm} is the result of
a LLA pQCD calculation~\cite{jhep:103:45}
based on the interaction of the proton with $q\bar{q}$ dipoles
with small transverse size
via two-gluon exchange. The model uses the CTEQ4L gluon PDF~\cite{cteq4l} 
evolved using skewed evolution.
This calculation is sensitive to the value of $\lambda$, a scaling parameter
that relates the transverse size of the dipole to the four-momentum
scales in the interaction cross section. 
The curve shown uses $\lambda$ =4, which gives a dependence on $W$ 
that is less steep than for $\lambda$ = 10, which was
favoured by studies of the proton structure function, $F_2$.

These predictions qualitatively describe the steep rise
of the cross section with energy.
At $\Wgp$ = 250 GeV,
the gluon density is being probed in these models~\cite{jhep:103:45}
at $x \sim 10^{-4}$, outside the range in which
it is well constrained by global PDF analyses; 
the results are, therefore, sensitive to the PDF used.
However, no discrimination between the gluon PDF 
can be made from the curves shown in Fig.~\ref{fig:cross-mrtcdm}
due to the large theoretical uncertainties
from higher-twist contributions
and the QCD scales due to missing higher-order terms.
In addition, skewed parton distributions~\cite{xji} are, as yet, 
relatively poorly constrained since the proton structure function $F_2$ 
is not very sensitive to them.

Also shown in Fig.~\ref{fig:cross-mrtcdm} is the result
(dot-dashed curve) of a calculation~\cite{soares00} 
based on a dipole model~\cite{pr:d59:014017,*pr:d60:114023}. 
The $J/\psi$ wave-function was assumed to be Gaussian 
in both the transverse and longitudinal momenta of the quarks.
The normalisation was fixed from the $b_0$ value reported in the next section.
The $W$ dependence of the model is in reasonable agreement with the 
present data.
%
\section{Differential cross-section 
$d\sigma /dt$ and the Pomeron trajectory \label{sec:traj}}
\begin{figure}[htbp!]
\epsfxsize=16cm
\centerline{\epsffile{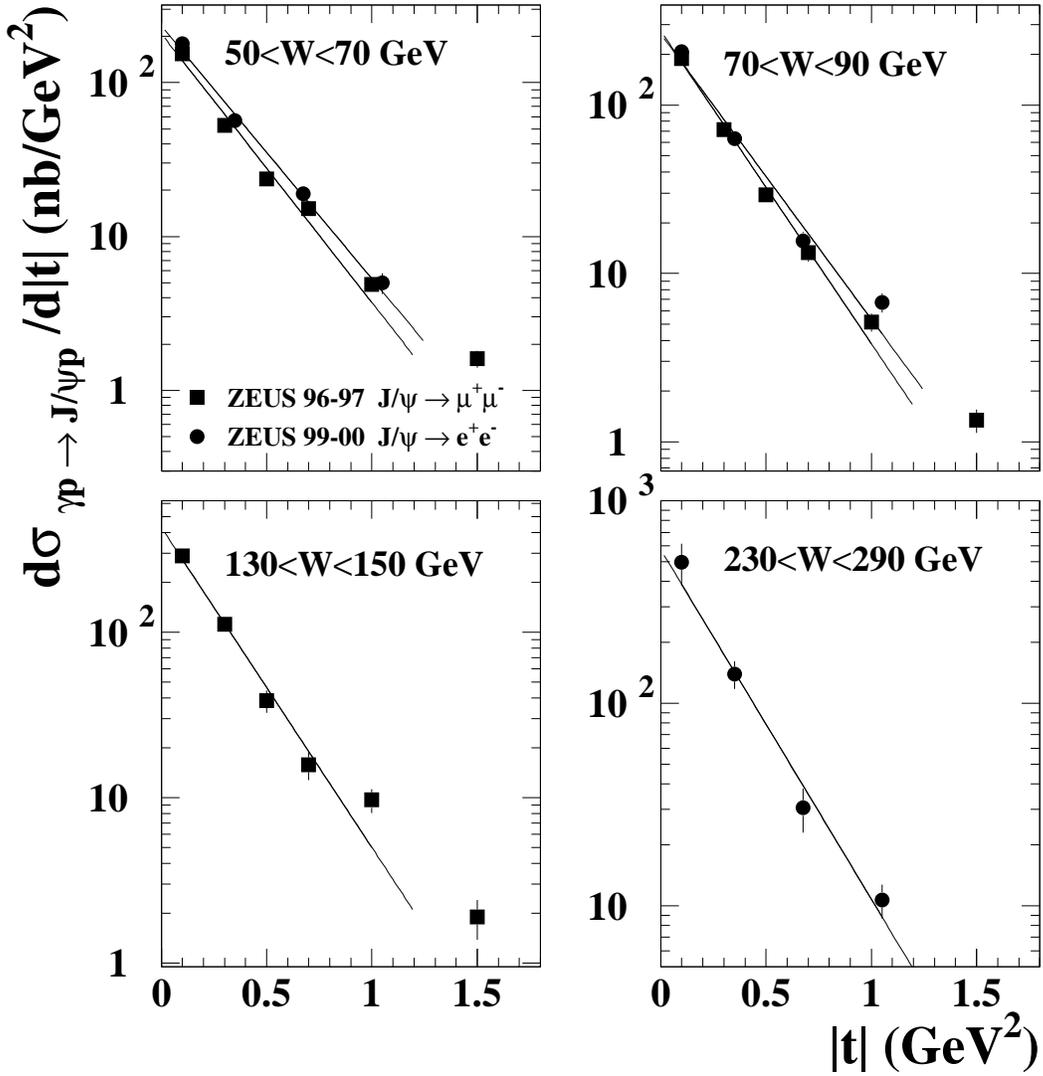}}
\caption{
The differential cross-section $d\sigma_{\gamma p \rightarrow J/\psi p}/dt$ 
for exclusive $J/\psi$ photoproduction for representative bins of $\Wgp$
and for the decay channels,
$J/\psi \rightarrow \mumu$ (squares)
and $J/\psi \rightarrow e^+e^-$ (points).
The vertical bars indicate the statistical uncertainties only.
The full lines represent the results of a fit of the form 
$d\sigma/dt = d\sigma/dt|_{t=0} \cdot e^{bt}$ 
performed in the range $-t<1.2 \gev^2$ for the muon channel and
in the range $-t<1.25 \gev^2$ for the electron channel.
}
\label{fig:dsdt_wbins}
\end{figure}

\begin{figure}[htbp!]
\epsfxsize=16cm
\centerline{\epsffile{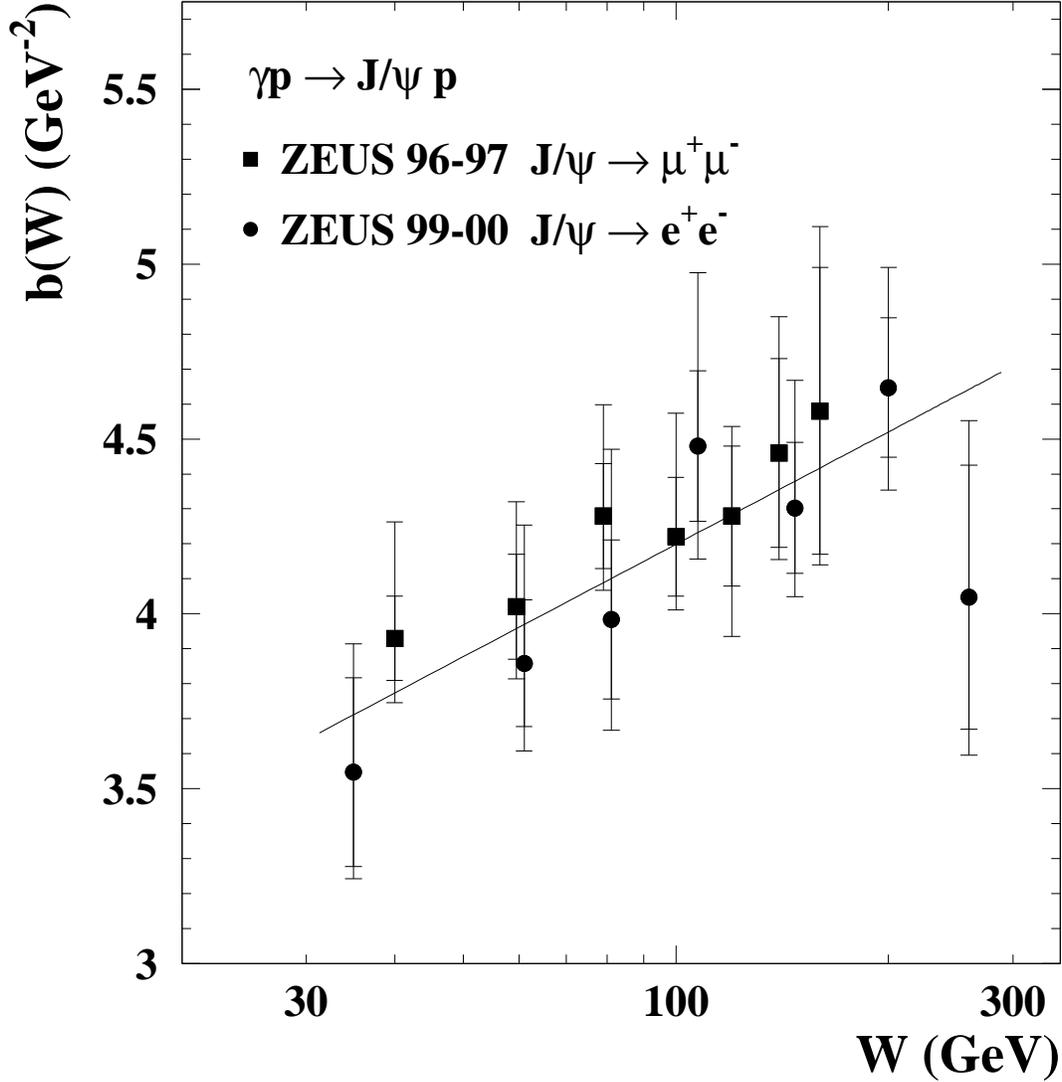}}
\caption{
Values of the slope, $b$, of the $t$ distribution, plotted
 as a function of $\Wgp$.
The line shows
the result of a fit of the form $b(W) =b(90\gev) +4 \cdot \alphappom \ln(W/90 \gev$).
}
\label{fig:bslope}
\end{figure}

%
The differential cross-section $d\sigma_{\gamma p \rightarrow J/\psi p}/dt$ 
was calculated, in bins of $\Wgp$, separately
for the muon and electron $J/\psi$
decay channels in the kinematic range $-t<1.8 \gev^2$ and
$-t<1.25 \gev^2$, respectively.
The results are shown in Fig.~\ref{fig:dsdt_wbins}
for four representative ranges of $W$. In each $W$ bin,
a fit of the form $d\sigma/dt = d\sigma/dt|_{t=0} \cdot e^{bt}$ 
was performed.
For the muon sample, the fit was performed in the restricted range 
$-t<1.2 \gev^2$,
where the uncertainty resulting from the subtraction of 
proton-dissociative events is small.
The results of the fits are given in Table~\ref{tab:xsection}.
The muon and electron analyses give consistent 
results for $d\sigma/dt|_{t=0}$ and $b$, as shown in
Figs.~\ref{fig:dsdt_wbins} and~\ref{fig:bslope}. 

The $b$ slope increases with $\Wgp$ and, 
in the geometrical picture of the interaction, 
is approximately equal to that expected 
from the size of the proton~\cite{hadhadt}, 
which suggests that the size of the 
$J/\psi$ is small compared to that of the proton. 

A value of $\alphappom$ was obtained by fitting
the $\Wgp$ dependence of $b$ to the function 
$b(W)=b_0+4\alphappom \cdot \ln(W/90 \gev)$,
according to Eq.~(\ref{eq:regge2}). The results for both
the muon and electron analyses, given in Table~\ref{tab:results},
are in good agreement. The systematic uncertainties were estimated 
by repeating the fit for each uncertainty not correlated in $\Wgp$
and adding the deviations from the nominal value in quadrature.
The result of the combined measurement, shown as the line in
Fig.~\ref{fig:bslope}, is
\[b_0 = 4.15 \pm 0.05(stat.)^{+0.30}_{-0.18}(syst.) \gev^{-2}\]
\[\alphappom = 0.116 \pm 0.026 (stat.)^{+0.010}_{-0.025}(syst.) \gev^{-2}.\]
The systematic uncertainties were computed from the combination of the
muon and electron analyses, taking into account the common systematic
uncertainties.

\begin{figure}[htbp!]
\epsfxsize=16cm
\centerline{\epsffile{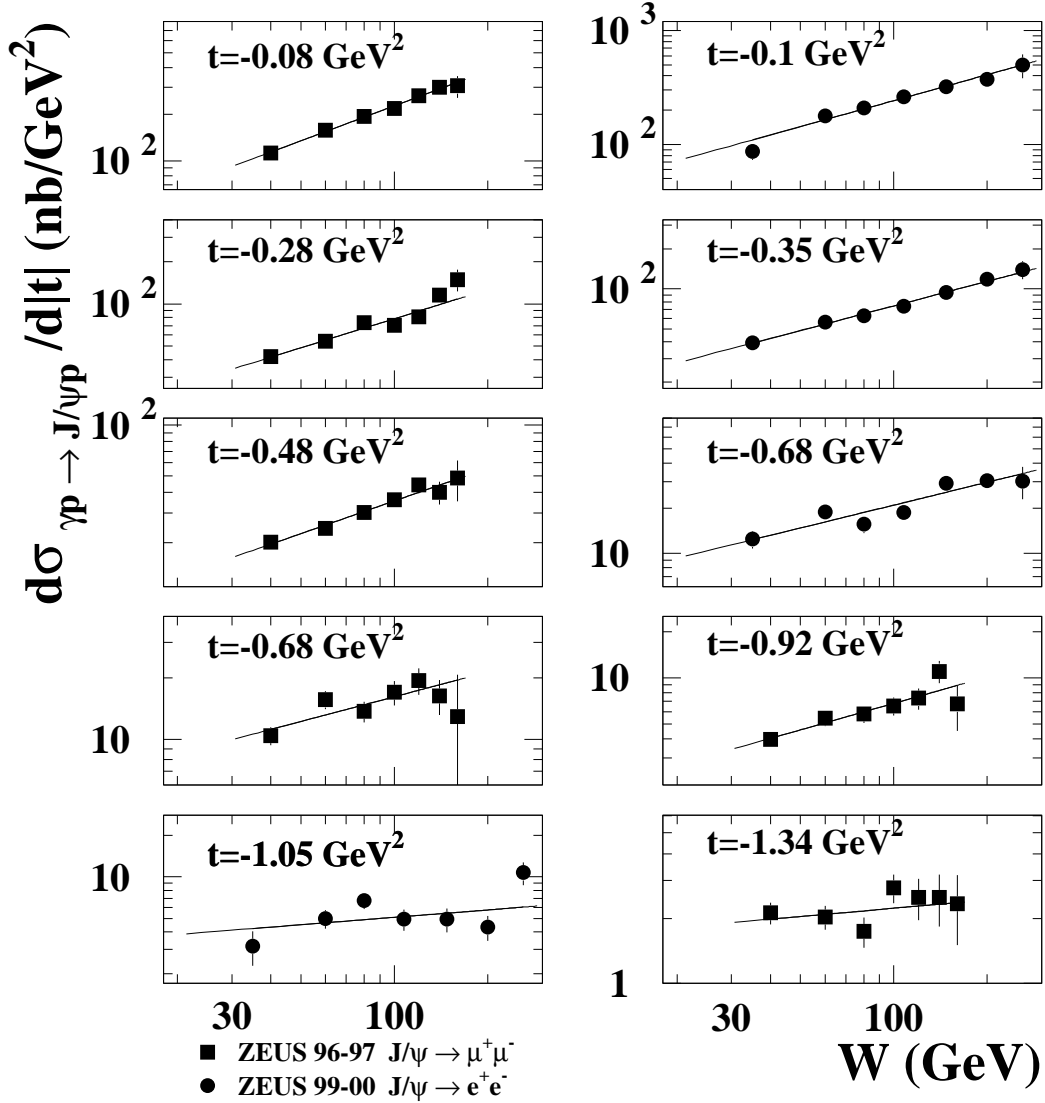}}
\caption{ 
The differential cross-section $d\sigma_{\gamma p \rightarrow J/\psi p}/dt$ 
as a function of $\Wgp$ at fixed $t$ values.
Only the statistical uncertainties are shown. 
The lines correspond to the results of fits of the form
d$\sigma_{\gamma p \rightarrow J/\psi p}/$d$t
\propto W^{4\cdot[\alphapom(t)-1]}$. 
}
\label{fig:dsdt-tbins}
\end{figure}

\begin{figure}[htbp!]
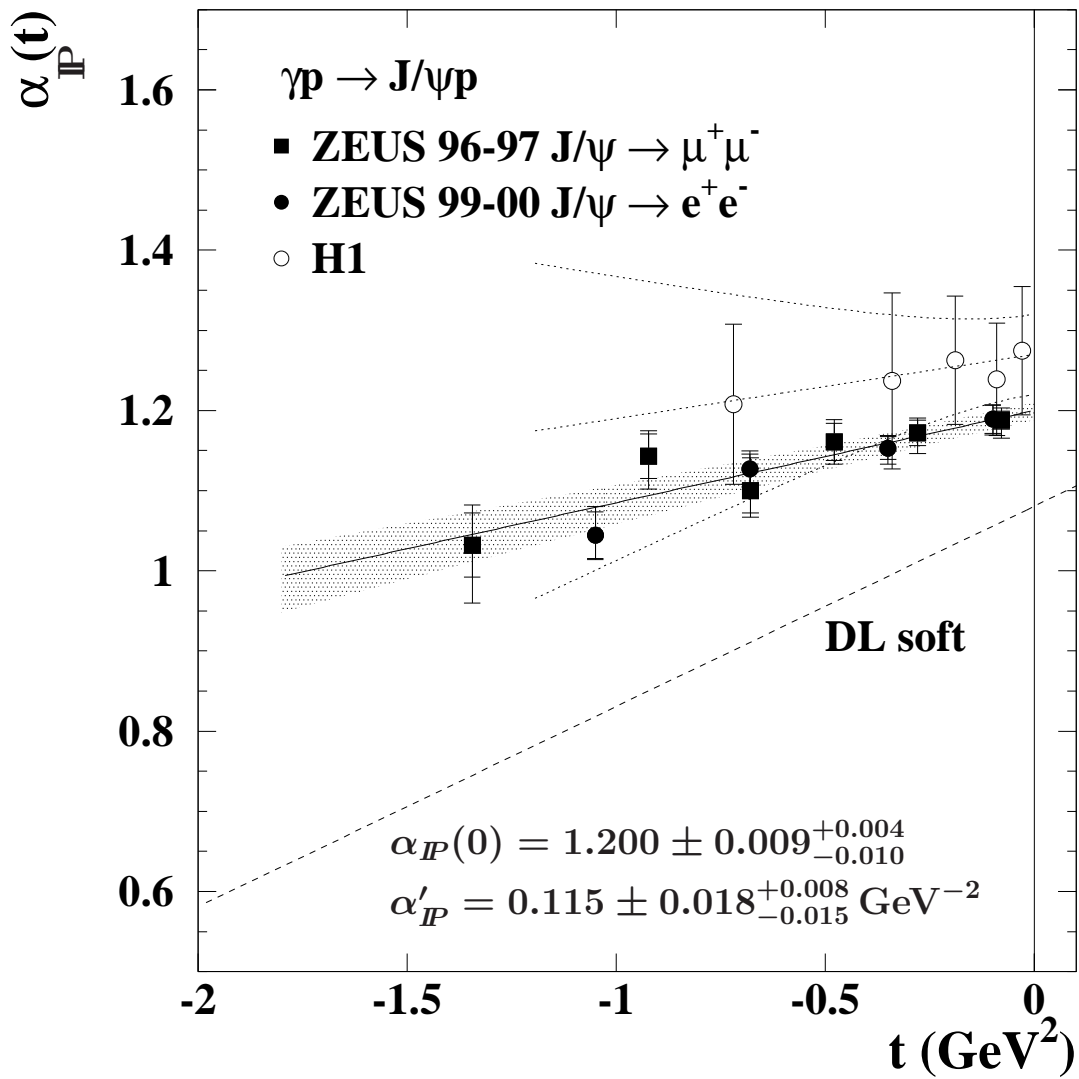

\psfigadd{fig.9}{16cm}{16cm}{
\Text(870,330)[]{\large\bf 
                $\boldsymbol{\alphapom(0) = 1.200\pm 0.009^{+0.004}_{-0.010}}$}
\Text(920,250)[]{\large\bf
                $\boldsymbol
{\alphappom = 0.115\pm 0.018^{+0.008}_{-0.015} \gev^{-2}}$}
\Text(113,1370)[]{\large\bf \begin{sideways} \pom \end{sideways} }
                                          }
\caption{ Pomeron trajectory as a function of $t$ 
as obtained in the two leptonic decay channels,
$J/\psi \rightarrow \mumu$ and $J/\psi \rightarrow e^+e^-$.
The inner bars indicate the  statistical uncertainties; 
the outer bars are the statistical
and systematic uncertainties added in quadrature. The results from the
H1 Collaboration~\pcite{H1001} are also shown.
The solid and dotted lines are the result of linear fits to the 
ZEUS and H1 data, respectively.
The one standard deviation contour is indicated for the 
ZEUS (shaded area) and H1 (dotted lines) measurements.
The dashed line shows the  DL soft-Pomeron trajectory~\pcite{DLsoft}.
}
\label{fig:alpha}
\end{figure}

The Pomeron trajectory was determined directly by 
measuring the variation of the $\Wgp$ dependence 
of the elastic cross section at fixed $t$, as parameterised
in Eq. (\ref{eq:regge1}). This method is
insensitive to  the proton-dissociative background,
since the latter was measured to be independent of $\Wgp$, 
as described in Section~\ref{background}.
In Fig.~\ref{fig:dsdt-tbins}, 
the measurements of $d\sigma/dt$ used in this determination 
of $\alphapom(t)$ are presented as a function of $\Wgp$
for fixed $t$;
the line in each plot is the result of a fit of the form
$d\sigma/dt \propto W^{4 \cdot[\alphapom(t)-1]}$. 
The resulting values of $\alphapom(t)$, given in Table~\ref{tab:alpha},
are shown in Fig.~\ref{fig:alpha}, 
with the published H1 results~\cite{H1001}, as a function of $t$.
They were fitted to the linear form
$\alphapom(t)=\alphapom(0) + \alphappom t$. The separate fits from the
muon and electron analyses are given in Table~\ref{tab:results} and are
in good agreement.
The combined measurement gives
\[\alphapom(0) = 1.200 \pm 0.009(stat.)^{+0.004}_{-0.010}(syst.)\]
and
\[\alphappom= 0.115 \pm 0.018(stat.)^{+0.008}_{-0.015}(syst.) \gev^{-2}.
\]
The systematic uncertainties were computed from the combination of the
muon and electron analyses, taking into account the common systematic
uncertainties.
The result of the fit was stable with respect to changes in the $t$ range
used for the fit.
The slope, $\alphappom$, measured with the present data, is not
consistent with zero and therefore indicates a small shrinkage.
The increase with $\Wgp$ of the cross section and of the $b$ slope,
parameterised by $\alphapom(0)$ and $\alphappom$, respectively, 
are in agreement 
with pQCD-based models~\cite{jetp:70:155,jhep:103:45}. 
The values of $\alphapom(0)$ and $\delta$ are compatible, 
after taking account of the 
measured value of $\alphappom$.
The soft-Pomeron trajectory 
$\alphapom =1.08 + 0.25 \cdot t$~\cite{DLsoft,pr:d10:170}  
is inconsistent with the present data. 
However, the contribution of the soft Pomeron plus a hard Pomeron
as proposed by Donnachie and Landshoff~\cite{DLhard} may well be 
able to describe the data.

\section{Conclusions}
The exclusive photoproduction of $J/\psi$ mesons has been
studied at HERA with the ZEUS detector 
in the kinematic range $20<\Wgp<290 \gev$
using both the muon and the electron decay channels.
The $J/\psi$ spin-density matrix elements, \rzfzz\ and \rzfpm\ ,
have been measured; their values are consistent, 
within two standard deviations, with the hypothesis of 
$s$-channel helicity conservation.

The $\gamma\, p\, \rightarrow\, J/\psi\, p$ cross section 
exhibits a strong dependence on $\Wgp$, which can be parameterised
 by a power-like
dependence of the type 
$W^\delta$, with $\delta=0.69 \pm0.02 (stat.) \pm 0.03 (syst.)$.
This behaviour is described by pQCD-based models and can be
understood as due to the increase of the  gluon density in the
proton for decreasing values of the parton fractional momentum.

The differential cross-section $d\sigma_{\gamma p \rightarrow J/\psi p}/dt$
has been measured 
as a function of $\Wgp$ for $|t|<1.8 \gev^2$.
It can be described by an exponential function in $t$, 
with a slope 
$b= 4.15 \pm 0.05(stat.)^{+0.30}_{-0.18}(syst.) \gev^{-2}$ at $W=90 \gev$,
which increases logarithmically with $W$.

The parameters of the Pomeron trajectory, $\alphapom(0)$ and $\alphappom$, 
have been determined from the $\Wgp$ and $t$
dependence of $d\sigma_{\gamma p \rightarrow J/\psi p}/dt$.
The intercept is 
$\alphapom(0)$ = $1.200\pm 0.009(stat.)$ $^{+0.004}_{-0.010}(syst.)$
and the slope is
$\alphappom= 0.115 \pm 0.018(stat.)^{+0.008}_{-0.015}(syst.) \gev^{-2}$.
These values are inconsistent with those expected 
from the exchange of a soft Pomeron.
The data indicate that $\alphappom$ is different from zero but smaller
by a factor of two than the value measured in soft hadronic interactions.

Clearly therefore, the description of $J/\psi$ production
lies within the realm of perturbative QCD. A quantitative description
comparable to the precision of the current data requires
further theoretical progress.
\section*{Acknowledgments}
We thank the DESY directorate for their strong support and
encouragement, and the HERA machine group for their diligent efforts.
We are grateful for the support of the DESY computing and network
services. The design, construction and installation of the ZEUS
detector have been made possible by the ingenuity and effort of many
people from DESY and home institutes who are not listed as authors. 
It is a pleasure to thank M. McDermott and T. Teubner 
for providing us with their model predictions and E. Levin for useful
discussions.


\newlength{\Tl}\settowidth{\Tl}{1.28$\pm  0.16^{+ 0.16}_{- 0.16}$}
%
\begin{table}[hP]
\begin{sideways}\begin{minipage}[b]
{\textheight}
\vspace{-0.5cm}    
\begin{center}
\begin{tabular}{||c|c|c|c|c|c|c|c|c|c||}
\hline

$W$ (GeV) & Mode & $N_{obs}$ & $N_{non-res}$ & $N_{\psi(2S)}$ 
& ${\cal A}$ & $\Phi$ & $\sigma$ (nb)
& $\frac{d\sigma}{dt}|_{t=0}$ ($\nb/\Gev^2$) & $b$ (GeV$^{-2}$)\\

\hline\hline
 20-30 & $\mumu$ & 139 & 36  & 7 & 0.0314  
         & 0.04520 & 32.6$\pm$ 5.4 $\pm$ 5.2 &\multicolumn{2}{c||}{}\\
\hline
 30-50 & $\mumu$ & 1883 &  207  &  12 & 0.260  
         & 0.05376 & 41.5$\pm 1.1 \pm 3.3$
         & $152 \pm  7 ^{+16}_{-9}$ & $3.93\pm 0.12 ^{+0.31}_{-0.14}$\\
 50-70 & $\mumu$ & 1512 & 54 &  19 & 0.281  
         & 0.03195 &  55.8$\pm 1.5 \pm 4.6$
         & $208 \pm 11 ^{+24}_{-16}$& $4.02\pm 0.15 ^{+0.26}_{-0.14}$\\ 
 70-90 & $\mumu$ &1299  & 62  & 16  & 0.294  
         & 0.02176 &  66.6$\pm 2.0 \pm 7.0$
         & $276 \pm 15 ^{+40}_{-29}$& $4.27\pm 0.15 ^{+0.28}_{-0.15}$\\ 
 90-110& $\mumu$ &1142 &  54 &  15 & 0.322  
         & 0.01585 &  73.4$\pm 2.3 \pm 6.0$
         & $295 \pm 18 ^{+36}_{-31}$& $4.22\pm 0.17 ^{+0.31}_{-0.12}$\\
110-130& $\mumu$ &842 &  45 &  5 & 0.265  
         & 0.01202 &  86.7 $\pm 3.2 \pm 6.5$
         & $356 \pm 25 ^{+35}_{-50}$& $4.28\pm 0.20 ^{+0.16}_{-0.28}$\\
130-150& $\mumu$ &541 &  48 &  2 & 0.176  
         & 0.009331& 104 $\pm 5 \pm 11 $
         & $430 \pm 42 ^{+60}_{-42}$& $4.46\pm 0.27 ^{+0.28}_{-0.14}$\\ 
150-170& $\mumu$ &171 &  29  & 6 & 0.0586 
         & 0.007367& 110 $\pm 11 \pm 12$
         & $488 \pm 84 ^{+80}_{-35}$ & $4.58\pm 0.41 ^{+0.33}_{-0.16}$\\ 
\hline
 20-35 & $e^+e^-$ & 982 & 216 & 16 &0.087 &  0.06408 &   33.6$\pm   1.6^{+  2.2}_{-  2.4}$ 
& \multirow{2}{\Tl}{128$\pm  16^{+ 16}_{- 16}$}
& \multirow{2}{\Tl}{3.55$\pm  0.27^{+ 0.25}_{- 0.14}$}\\
 35-50 & $e^+e^-$ &2681 & 881 & 49 &0.270 &  0.03730 &   43.8$\pm   2.0^{+  2.9}_{-  2.8}$ 
&&\\

\hline
 50-60 & $e^+e^-$ &1978 & 372 & 30 &0.386 &  0.01791 &   57.2$\pm   1.8^{+  3.6}_{-  3.5}$ 
& \multirow{2}{\Tl}{228$\pm  17^{+ 35}_{- 19}$} 
& \multirow{2}{\Tl}{3.86$\pm  0.18^{+ 0.35}_{- 0.17}$ }\\
 60-70 & $e^+e^-$ &1821 & 408 & 35&0.383 &  0.01447 &   62.5$\pm   2.3^{+  4.1}_{-  3.9}$ &&\\

\hline
 70-80 & $e^+e^-$ &1577 & 326 & 30&0.371 &  0.01200 &   68.9$\pm   2.6^{+  4.5}_{-  4.5}$ 
&\multirow{2}{\Tl}{258$\pm  26^{+ 45}_{- 25}$} 
&\multirow{2}{\Tl}{3.98$\pm  0.23^{+ 0.43}_{- 0.22}$} \\
 80-90 & $e^+e^-$ &1420 & 306 & 30&0.373 &  0.01013 &   72.1$\pm   2.9^{+  4.9}_{-  4.5}$ &&\\

\hline
 90-110 & $e^+e^-$ &2499 & 445 & 35&0.382 &  0.01620 &   81.9$\pm   2.3^{+  4.9}_{-  4.8}$
&\multirow{2}{\Tl}{360$\pm  31^{+ 58}_{- 34}$} 
&\multirow{2}{\Tl}{4.48$\pm  0.22^{+ 0.45}_{- 0.24}$} \\
110-125 & $e^+e^-$ &1604 & 235 & 21&0.371 &  0.00954 &   95.7$\pm   3.2^{+  5.4}_{-  5.4}$ &&\\

\hline
125-140 & $e^+e^-$ &1470 & 221 & 14&0.378 &  0.00790 &  103.9$\pm   3.6^{+  6.0}_{-  5.8}$
&\multirow{2}{\Tl}{439$\pm  38^{+ 57}_{- 35}$} 
&\multirow{2}{\Tl}{4.30$\pm  0.19^{+ 0.31}_{- 0.17}$} \\
140-170 & $e^+e^-$ &2303 & 393 & 46&0.334 &  0.01218 &  115.0$\pm   3.3^{+  7.7}_{-  6.7}$ &&\\

\hline
170-200 & $e^+e^-$ &1362 & 212 & 23 &0.250 &  0.00877 &  129.1$\pm   4.7^{+  8.1}_{-  7.7}$
&\multirow{2}{\Tl}{578$\pm  57^{+ 72}_{- 76}$} 
&\multirow{2}{\Tl}{4.65$\pm  0.20^{+ 0.28}_{- 0.21}$} \\
200-230 & $e^+e^-$ &1161 & 261 & 19 &0.243 &  0.00643 &  141.7$\pm   6.1^{+  8.8}_{-  8.7}$ &&\\

\hline
230-260 & $e^+e^-$ &1208 & 442 & 13 &0.280 &  0.00482 &  140.3$\pm   7.4^{+ 15.1}_{-  9.9}$
&\multirow{2}{\Tl}{557$\pm 120^{+ 88}_{- 93}$} 
&\multirow{2}{\Tl}{4.05$\pm  0.38^{+ 0.33}_{- 0.25}$} \\
260-290 & $e^+e^-$ &1490 & 785 & 18 &0.248 &  0.00369 &  189$\pm  13^{+ 15}_{- 26}$ &&\\

\hline
\end{tabular}
\hspace*{-8.5cm}\begin{minipage}[b]{27.5cm}
\setlength{\captionwidth}{23.cm}
\isucaption{
Measurements, in different ranges of $W$,
of the total $J/\psi$ photoproduction cross section,
of the differential cross section 
extrapolated to $t=0$ and of the 
slope parameter $b$ of the exponential $t$ dependence.
The first uncertainties are statistical and the second are systematic.
$N_{obs}$ is the number of events in the signal mass region,
$N_{non-res}$ is the estimated non-resonant background, 
$N_{\psi(2S)}$ is the number of $\psi(2S)$ events
in the $J/\psi$ mass region and
${\cal A}$ is the acceptance.
The effective photon flux, $\Phi$, is used to compute the $\gamma p$ 
cross section from the $ep$ cross section. 
\label{tab:xsection}
}
\end{minipage}

\end{center}
\end{minipage}\end{sideways}
\end{table}

%
\begin{table}[hP]
\begin{center}
\begin{tabular}{||c|c|c|c||}
\hline
Quantity & $J/\psi \rightarrow \mu^+\mu^-$ & $J/\psi \rightarrow e^+ e^-$ & 
Method \\
\hline\hline
$\delta$ & $0.67 \pm 0.03 \pm 0.05 $ & $0.695 \pm 0.021 \pm 0.028$ & Fits to\\ 
$\Wgp$ range & $30<\Wgp < 170$ GeV & $35<\Wgp<290$ GeV & 
$\sigma \propto (\Wgp/90 {\rm GeV})^{\delta}$ \\
\hline
$b_0$ ($\Gev^{-2}$) 
                &$4.23 \pm 0.07^{+0.10}_{-0.12}$$^{+0.085}_{-0.051}$ &
                 $4.11 \pm 0.08^{+0.08}_{-0.09}$$^{+0.33}_{-0.16}$   & Fits to \\
$\alphappom$ ($\Gev^{-2}$)
                &$0.098 \pm 0.037 \pm 0.040 \pm 0.001$      &
                 $0.128 \pm 0.037^{+0.008}_{-0.025}\pm 0.005$ &
                Eq.~(\ref{eq:regge2}) \\
\hline
$\alphapom (0)$ &$1.198 \pm 0.011 \pm 0.015$    &
                 $1.204 \pm 0.016^{+0.004}_{-0.013}$     & Fits to \\
$\alphappom$ ($\Gev^{-2}$)
                &$0.099 \pm 0.023 \pm 0.020$    &
                 $0.136 \pm 0.031^{+0.008}_{-0.020}$ &     
                Eq.~(\ref{eq:regge1}) \\
$t$ range       &  $-t < 1.8$ GeV$^2$ & $-t < 1.25$ GeV$^2$ & \\
\hline
\end{tabular}    
\caption{Measurements of $\delta$, $b_0$, $\alphappom$ and $\alphapom (0)$
obtained separately from the muon and electron decay channels.
The first uncertainties are statistical and the second are systematic.
Where given, the third refers to the modelling of the proton-dissociative
subtraction. 
The last column indicates how the values were determined.
}
\label{tab:results}
\end{center}
\end{table}

%
\begin{table}[hP]
\begin{center}
\begin{tabular}{||c|c|c||}
\hline
$t$ (GeV$^2$) & Mode &  $\alphapom (t)$ \\
\hline\hline
-0.079 & $\mumu$ & $1.188\pm 0.011 ^{+0.010}_{-0.020} $\\ 
-0.28  & $\mumu$ & $1.172\pm 0.016 ^{+0.010}_{-0.020} $\\ 
-0.48  & $\mumu$ & $1.161\pm 0.023 ^{+0.015}_{-0.016} $\\ 
-0.68  & $\mumu$ & $1.100\pm 0.028 ^{+0.030}_{-0.017} $\\ 
-0.92  & $\mumu$ & $1.143\pm 0.028 ^{+0.015}_{-0.030} $\\ 
-1.34  & $\mumu$ & $1.032\pm 0.040 ^{+0.030}_{-0.060} $\\ 
   -0.10 & $e^+e^-$ &   1.189$\pm    0.018^{+   0.005}_{-   0.009}$\\
   -0.35 & $e^+e^-$ &   1.153$\pm    0.014^{+   0.006}_{-   0.014}$\\
   -0.68 & $e^+e^-$ &   1.127$\pm    0.019^{+   0.012}_{-   0.011}$\\
   -1.05 & $e^+e^-$ &   1.044$\pm    0.029^{+   0.021}_{-   0.009}$\\
 \hline
\end{tabular}    
\caption{Values of $\alphapom (t)$ 
obtained from fits to the function $d\sigma/dt \propto 
W^{4\cdot[\alphapom (t)-1]}$.
The first uncertainty is statistical and the second is the systematic.
}
\label{tab:alpha}
\end{center}
\end{table}

%
%
\end{document}